\documentclass{aa}  

\usepackage{graphicx}
\usepackage{txfonts}
\usepackage{caption}
\usepackage[dvipsnames,table]{xcolor}
\usepackage{float}
\usepackage{placeins}
\usepackage{multirow}

\newcommand{\ero}{eROSITA}

\newcommand{\xmm}{\textit{XMM-Newton}\,}

\begin{document}

   \title{The SRG/eROSITA All-Sky Survey}
   \subtitle{View of the Virgo Cluster}

   \titlerunning{eROSITA view of the Virgo Cluster}

   \authorrunning{H. McCall et al.}

   \author{Hannah McCall
          \inst{1, 2},
        Thomas H. Reiprich \inst{1},
        Angie Veronica \inst{1},
        Florian Pacaud \inst{1},
        Jeremy Sanders \inst{3},
        Henrik W. Edler\inst{4},
        Marcus Br\"uggen\inst{4},
        Esra Bulbul\inst{3},
        Francesco de Gasperin\inst{4,5},
        Efrain Gatuzz\inst{3},
        Ang Liu\inst{3},
        Andrea Merloni\inst{3},
        Konstantinos Migkas\inst{1,6,7},
        \and
        Xiaoyuan Zhang\inst{3}
          }

   \institute{
             Argelander-Institut f\"ur Astronomie (AIfA), Universit\"at Bonn, Auf dem H\"ugel 71, 53121 Bonn, Germany\\
             \email{hannah.mccall@uni-bonn.de}
        \and
            Department of Astronomy and Astrophysics, University of
            Chicago, Chicago, IL 60637, USA
        \and
            Max-Planck-Institut f\"ur extraterrestrische Physik, Giessenbachstrasse, 85748 Garching, Germany
        \and
            Hamburger Sternwarte, University of Hamburg, Gojenbergsweg 112, D-21029, Hamburg, Germany
        \and
            INAF - Istituto di Radioastronomia, via P. Gobetti 101, 40129, Bologna, Italy
        \and
            Leiden Observatory, Leiden University, PO Box 9513, NL-2300 RA Leiden, The Netherlands
        \and
            SRON Netherlands Institute for Space Research, Niels Bohrweg 4, NL-2333 CA Leiden, The Netherlands
}

   \date{Received January 30, 2024; accepted May 29, 2024}

 
  \abstract
   {As the closest galaxy cluster, the Virgo Cluster is an exemplary environment for the study of the large-scale filamentary structure and physical effects that are present in cluster outskirts but absent from the more easily studied inner regions.}
   {Here, we present an analysis of the SRG/eROSITA data from five all-sky surveys.}
   {eROSITA allows us to resolve the entire Virgo cluster and its outskirts on scales between 1 kpc and 3 Mpc, covering a total area on the sky of about 25$^\circ$ by 25$^\circ$. 
   We utilized image manipulation techniques and surface brightness profiles to search for extended emission, surface brightness edges, and features in the outskirts. We employed a method of comparing mean and median profiles to measure gas clumping out to and beyond the virial radius.
   }
   {Surface brightness analysis of the cluster and individual sectors of the cluster reveal the full extent of previously identified cold fronts to the north and south. The emissivity bias due to gas clumping, which we quantify over three orders of magnitude in the radial range, is found to be mild, consistent with previous findings. We find uniform clumping measurements in all directions, with no enhancements along candidate filaments. We find an estimated virial gas mass of $M_{\mathrm{gas},r<r_{200}} = (1.98 \pm 0.70) \times 10^{13}$\,M$_\odot$. Through imaging analysis we detect the presence of extended emission spanning 320\,kpc to the southwest of M49. The extension has a significance of 3.15$\sigma$ and is coincident with radio emission detected with LOFAR, which could be evidence of an accretion shock or turbulent reacceleration as the M49 group or background W' cloud falls into the cluster and interacts with the intracluster medium. }
   {}


   \maketitle
%

\section{Introduction}

    The Virgo Cluster is the closest ($\sim16.1$\,Mpc,  \citealt[][]{Tonry_2001}) galaxy cluster and one of the best studied. In the X-ray band, in which the strongest emission processes are those of thermal bremsstrahlung and line emission in the hot intracluster medium (ICM), the interior of Virgo and its brightest cluster galaxy (BCG) M87 have been the subject of much previous study \citep{Bo1994, Bo_1995, Schindler_1998, Churazov_2001, Young_2002, Forman_2005, Werner_2006, Forman_2007, Simionescu_2008, Million_2011, Arevalo_2016, Gatuzz_2022, Gatuzz_2023}. Other works have characterized portions of the outskirts of Virgo \citep{Urban2011, Simionescu_2017, Mirakhor_2021}. The outskirts of clusters, which \citet[][]{Reiprich2013} defines as between $r_{500}$ and $3r_{200}$, can be a location of physical effects that are no longer observable in the more relaxed central regions. These include breakdown of hydrostatic, thermal, equipartition, and ionization equilibrium states, as well as structure formation effects, such as clumpy gas distribution or accretion shocks (see \citealt{Reiprich2013}, \citealt{Walker_2019} for a review of these processes in cluster outskirts). Limited by high particle backgrounds, however, previous X-ray instruments have been largely unable to observe significant emission in the outskirts, or else have had too narrow a field of view (FOV) to cover the entirety of the closest clusters.

    The extended Roentgen Survey and Imaging Telescope Array (\ero) was launched aboard the Spektrum Roentgen Gamma (SRG) mission \citep{Sunyaev_2021} in 2019 with a main science goal of detecting the hot ICM of galaxy clusters and groups to enhance the study of cosmic structure evolution \citep[][]{Predehl_2021}. eROSITA is expected to detect $10^5$ galaxy clusters and groups through its all-sky survey \citep{Merloni2012, Liu_2022, Bulbul_2022}. eROSITA's combination of soft-band sensitivity and a wide FOV makes it the ideal instrument for the study of faint emission in cluster outskirts that may have previously gone undetected. Other current generation X-ray instruments, such as \textit{Chandra} and \textit{XMM-Newton}, have a relatively small FOV and are thus inefficient in mapping large volumes of the Universe; only through mosaics have they been able to study out to and beyond $r_{200}$. The only imaging X-ray all-sky survey to precede eROSITA  is ROSAT, which was performed over six months in 1990. By the completion of the eight all-sky surveys planned for \ero, its survey results will be 25 times more sensitive than ROSAT in the soft band (0.2-2.3\,keV, \citealt[][]{Merloni2012, Predehl_2021}).

    The Virgo Cluster's distance, at which 1 arcminute corresponds to 4.65\,kpc, provides the opportunity to study the ICM on smaller scales than is possible with any other cluster. Recent studies with other X-ray instruments have taken advantage of Virgo's distance, modest physical size, and dynamism to better analyze its outskirts. These include a study by \citet{Urban2011}, which uses a series of thirteen \textit{XMM-Newton} pointings out to the virial radius in the north to characterize the outskirts with spectral analysis; observations with \textit{XMM-Newton} to study gas clumping by \citet{Mirakhor_2021}; and a Suzaku mosaic along four arms out to the virial radius, which \citet{Simionescu_2017} uses to analyze the thermodynamics of the outskirts. All of these studies were limited to analysis along a single radial line to $r_{200}$ due to the narrow FOV of the instruments used. 

    Here, we present an eROSITA imaging analysis of the Virgo Cluster produced from the cumulative data of its first four complete surveys and partial fifth survey. This is the first comprehensive X-ray imaging analysis of the cluster since ROSAT and it reveals significantly more detail. This article is organized as follows: in Sect.\,\ref{sec:data}, we describe the data reduction steps and analysis strategy. In Sect.\,\ref{sec:results}, we present what the imaging analysis and surface brightness analysis reveal about the large-scale structure of the cluster. We describe the results of deprojection analysis in Sect.\,\ref{sec:deprojection}. Section\,\ref{sec:M49} describes a region to the southwest of M49 where eROSITA soft-band X-ray emission is found to be coincident with LOFAR radio emission. We conclude with a summary 
    in Sect.\,\ref{sec:conclude}.

    Throughout this work, we assume a $\Lambda$CDM cosmology with $\Omega_m = 0.27$, $\Omega_{\Lambda} = 0.73$, and H$_0 = 70$\,km/s/Mpc. The distance to the Virgo Cluster is taken to be that of its central galaxy, M87, at 16.1\,Mpc \citep[][]{Tonry_2001}, or $z \approx 0.00377$. For consistency with other works \citep[][]{Urban2011, Simionescu_2017}, we consider $r_{200}$, the radius within which the mean density is 200 times the critical density of the Universe, to be the virial radius. The virial radius of the Virgo Cluster is taken to be 1.08\,Mpc \citep{Urban2011}, or a projected radius of $\sim3.9^{\circ}$. Other radii (i.e., $r_{500}$) are calculated with the relationships given in \citet[][]{Reiprich2013}. Abundances are given relative to the solar abundance of \citet{Asplund_2009}. Unless otherwise stated, all uncertainties are given at a 68 percent confidence level.

\section{Data reduction and analysis} \label{sec:data}

Observations of the Virgo Cluster from the first five eROSITA  all-sky surveys (eRASS:5) were used in this analysis, although the fifth survey only covers approximately half of the area (see \citealt{Merloni_2024} for all-sky survey information, sky tile definition, etc.). A comprehensive list of the 62 utilized sky tiles, which span an approximate area of $25^\circ$ by $25^\circ$, similar to the area covered by >2500 \xmm EPIC pointings, can be found in \autoref{app:Observations}. The processing configuration c020, extended Science Analysis Software (eSASS, \cite{Brunner_2022}) version \texttt{eSASSusers\_211214}, and \texttt{Heasoft} version 6.29 were used throughout the data reduction.

\subsection{Data preparation}

Upon retrieving the data, the first action was to merge the eRASS:5 event lists of all seven\footnote{Not all TMs could be used for every eRASS or tile. See \autoref{app:Observations} for a list of the excluded TMs.} telescope modules (TMs) across the full energy range 0.2-10\,keV for each tile using the eSASS task \texttt{evtool}. In this work, we use the notation that TM\,0 is the combination of all seven TMs. The parameters \texttt{pattern=15} and \texttt{flag=0xe00fff30} were applied to select single, double, triple, and quadruple patterns and remove bad pixels and strongly vignetted corners of CCDs. 
The eSASS task \texttt{flaregti} was then run with parameters \texttt{source\_size = 150}, which is in units of arcseconds and dictates the diameter of the extraction area for dynamic threshold calculation, \texttt{pimin=5000}, which is in units of electron volts and controls the lower energy range used, and \texttt{gridsize=26}, which determines the number of grid points per dimension for the dynamic threshold calculation. This task produces updated good time intervals (GTIs) that exclude times during which there is flare contamination. In addition to removing several short flares, \texttt{flaregti} aided in the removal of a flare that spanned the entire area of analyzed sky during eRASS\,2.

Following flare filtering, the eSASS task \texttt{radec2xy} was applied to recenter all tiles around a common center, R.A.\ 187.7042$^\circ$, Dec.\ +12.3911$^\circ$. Updated GTIs were applied using \texttt{evtool} with the parameters \texttt{gti}=``FLAREGTI" and \texttt{rebin}=124, which results in a pixel size of 6.2 arcseconds. This choice allowed the entire sky area to fit into the maximum eSASS image size. The combined event file was then split into its seven individual TMs for further image processing, with the ultimate goal of a cleaned and corrected TM\,0 image. The energy band for the analysis was restricted to 0.3-2.0\,keV for TMs with on-chip filter (TMs 1, 2, 3, 4, and 6; combined, they are referred to as TM\,8) and 0.8-2.0\,keV for TMs without on-chip filter (TMs 5 and 7; referred to as TM\,9). The reason for this differing lower limit is the optical light leak contamination of TM\,9, which is further detailed in \citet[][]{Predehl_2021}. Both vignetted and non-vignetted exposure maps were created for every image using the task \texttt{expmap}.

\subsection{Image corrections}

\begin{figure*}
	\centering
	\includegraphics[width=0.75\linewidth]{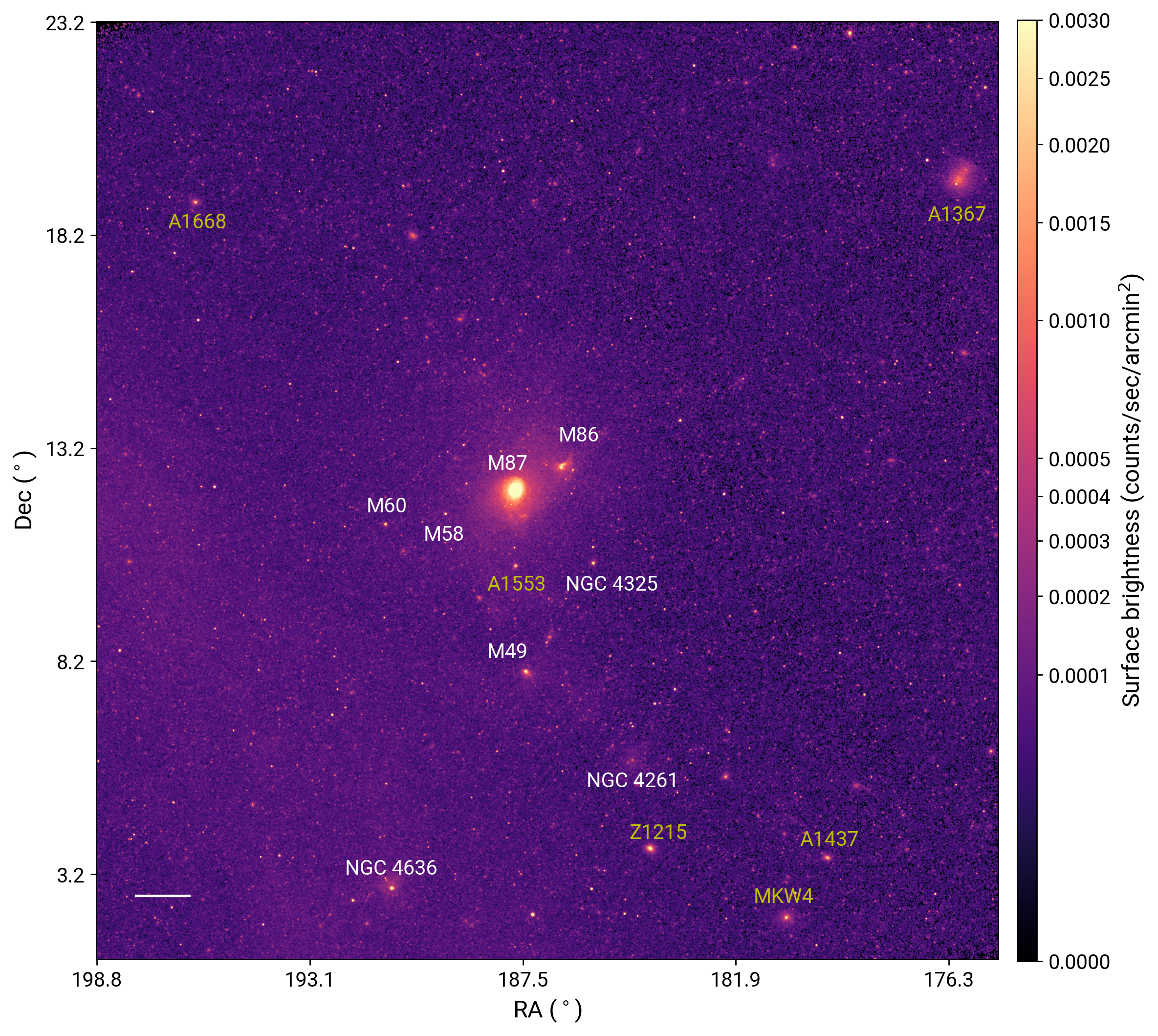}
	\caption{eROSITA TM\,0 particle-induced background subtracted, exposure corrected, absorption corrected count rate image in the $0.3-2.0$\,keV energy band. The image is displayed in logarithmic scaling with a Gaussian smoothing of radius 18 pixels, or 112 arcseconds. The white bar in the lower left spans 1 degree, or 279\,kpc. Prominent groups and galaxies within Virgo are labeled in white, while non-members of the cluster are labeled in yellow.}
	\label{fig:final_image}
\end{figure*}

In order to better isolate the ICM emission visible in the eROSITA image, it is necessary to account for the particle background, spatially variable absorption, and variable coverage of the field by the different TMs.

\subsubsection{Particle-induced background}

 The particle-induced background (PIB) subtraction steps are based on knowledge of the eROSITA Filter-Wheel-Closed (FWC) data \citep[][]{Freyberg_2020, Yeung_2023} and were performed as described in \citet[][]{Reiprich_2021}. Important TM-specific values for this calculation can be found in \autoref{tab:PIB}. This process results in PIB maps for each TM, which are then added to create a TM\,0 PIB map that can be subtracted from the photon image.

\subsubsection{Relative absorption correction}

The next step was to account for the spatially variable absorption, since $N_\mathrm{HI}$ varies across the large sky area and can potentially introduce a bias in soft emission detection. The $N_\mathrm{HI}$ map covering the sky area around the Virgo Cluster (see Fig.~\ref{fig:nh_map}) is based on the 21 cm HI4PI survey \citep[][]{Bekhti_2016}. Its values span $(0.71-5.95) \times 10^{20}\,\mathrm{cm}^{-2}$. Thus, in order to calculate the $N_\mathrm{HI}$ correction factor across the entire image, spectra for $525$ $N_\mathrm{HI}$ values in this range were simulated to quantify the effect of varying absorption on the X-ray spectra and overall count rates. The simulations were conducted in \texttt{xspec} version $12.12.0$ \citep{Arnaud_1996}. The \texttt{fakeit} command was used with the model

\begin{align}\label{eq:spectrum}
	apec_{1} + tbabs \times \left(apec_{2} + pow\right) ,
\end{align}
where $apec_1$ accounts for the X-ray emission from the unabsorbed Local Hot Bubble (LHB), $tbabs$ corresponds to the absorption along the line-of-sight, $apec_2$ accounts for the X-ray emission of the Milky Way Halo (MWH), and $pow$ is the power law component originating from unresolved active galactic nuclei (AGN). The temperatures, metal abundances, and redshifts of the LHB and MWH are fixed, along with the power law photon index. Due to the origin within the galaxy, both metal abundances are set to $1\, \mathrm{Z}_{\odot}$, while both redshifts are set to zero. All values can be found in \autoref{tab:nh}.

The simulations were run for TMs with and without on-chip filter. This required the use of their separate response files and soft energy bands. The result of the simulated spectra was count rates for every $N_\mathrm{HI}$ value. From these, a ratio of the count rate for a given $N_\mathrm{HI}$ to the count rate of the reference $N_\mathrm{HI}$ can be calculated. The reference $N_\mathrm{HI}$ is the median $N_\mathrm{HI}$ across the total sky area and in this instance was $3.33 \times 10^{20}\,\mathrm{cm}^{-2}$. The correction factor map is then acquired by distributing the ratio corresponding to a given $N_\mathrm{HI}$ value to its area on the sky. The correction maps for TM\,8 and TM\,9 are applied by multiplying them by the exposure maps. 

The variation of the correction factors is 18.6\% for TM\,8 and 7\% for TM\,9. Since absorption is weaker at higher energies, the smaller variation of TM\,9 is expected.

\subsubsection{Exposure correction}

Since TM\,8 and TM\,9 feature different soft responses and different energy bands to account for the light leak, it is necessary to apply a correction factor when combining all TMs and observations. This process, referred to as exposure correction and described in larger detail in \citet[][]{Reiprich_2021}, involves using the TM\,9-to-TM\,8 ratio of the $N_{HI}$-corrected, PIB-subtracted count rates. The correction factor for this analysis is $0.392$. Exposure maps prior to and following absorption and exposure correction can be seen in Fig.~\ref{fig:exp_maps}.

The final image is found by adding the individual TM event files together to create the TM\,0 photon image, subtracting the PIB map, and dividing by the corrected exposure map. It is shown in Fig.~\ref{fig:final_image}. Several prominent groups, galaxies, and clusters are labeled.

\section{Larger Scale Structure} \label{sec:results}

\subsection{X-ray images} \label{sec:regions}

\begin{figure*}
	\centering
	\includegraphics[width=0.8\linewidth]{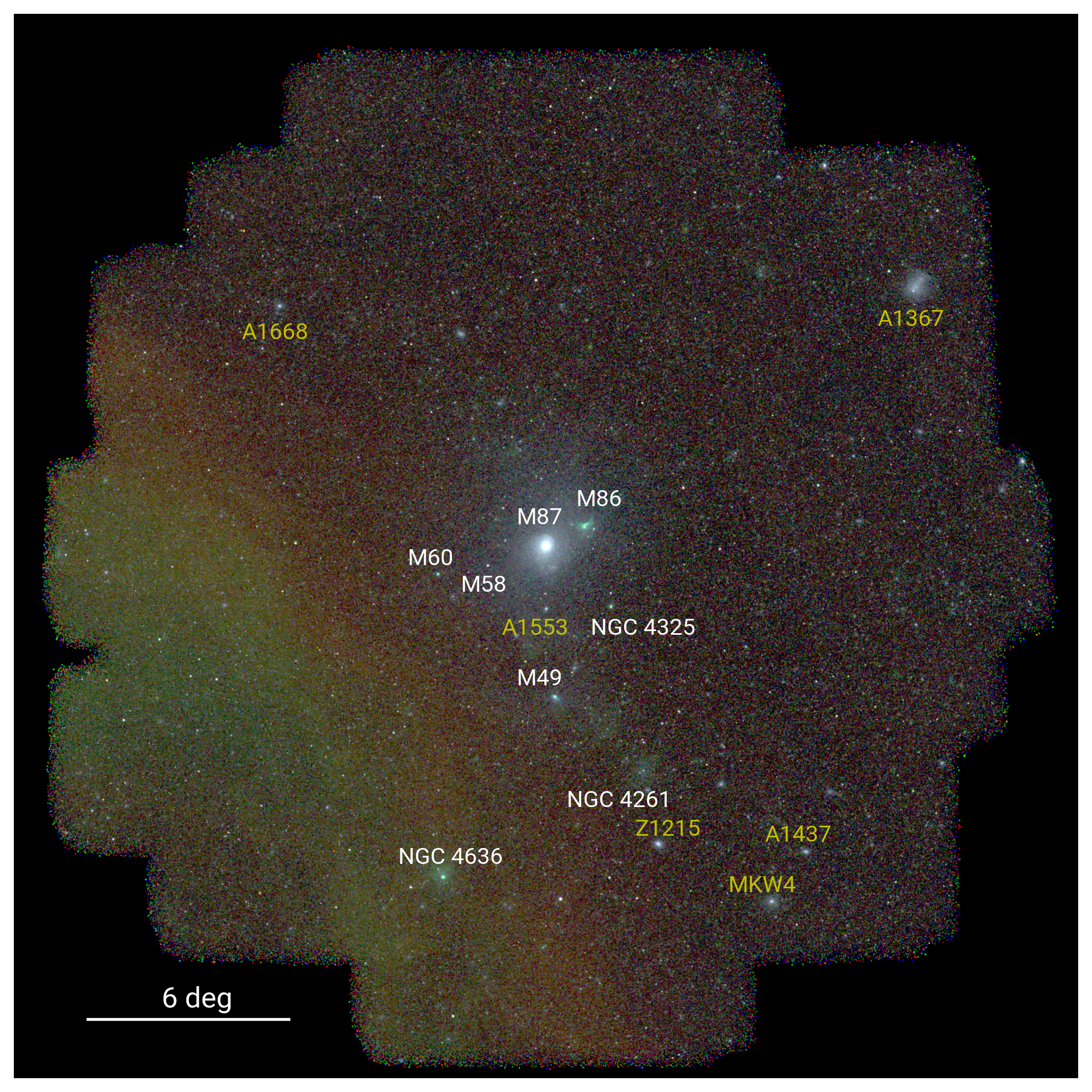}
	\caption{RGB map of the Virgo Cluster out to three times the virial radius of the cluster (red for $0.3-0.6$\,keV, green for $0.6-1.0$\,keV, and blue for $1.0-2.3$\,keV) is shown. The surface brightness for all channels has a dynamic range from $10^{-5}$ to $1.8 \times 10^{-3}$ counts/sec/arcmin$^2$. The map was rescaled by a factor of 5 in each dimension, but no smoothing was applied.}
	\label{fig:rgb_image}
\end{figure*}

An RGB image can be seen in Fig.~\ref{fig:rgb_image}. This was created following the steps described in Sect.~\ref{sec:data}, except that different energy ranges were used. The colors and corresponding bands are: red for $0.3-0.6$\,keV, green for $0.6-1.0$\,keV, and blue for $1.0-2.3$\,keV. We note that TM\,9 was not used to create the red band, and was only used in the range $0.8-1.0$\,keV for the green band. In the image, the diffuse ICM emission around M87 is visible in a hazy white, while many of the galaxies in Virgo, most obviously M86 and NGC 4636, appear green-blue. This implies that these regions have a harder spectrum than the surrounding medium, which could result from a higher temperature. Outside of the virial radius, extended white-green emission is also visible surrounding and to the southwest of M49. The point sources are likely a combination of foreground stars and background AGN. The variety of point source colors implies a plethora of spectral shapes. The northern eROSITA bubble \citep{Predehl_2020}, also known as the North Polar Spur, is bluer towards the Galactic center (outside the analyzed area to the southeast), and then fades to green and then red closer to the center of Virgo, indicating that the highest energies are closest to the Galactic center. This observation is in agreement with the assertion in \citet{Predehl_2020} that the bubbles are likely due to energy injections from the Galactic center. 

We can infer from the color of the eROSITA bubbles that the soft Virgo Cluster emission in the southeast is likely contaminated by bubble emission. Comparison of the surface brightness in the lower left and upper right of the image revealed that the bubble emission enhances the foreground by a factor of about two. In the blue band, however, there is almost no contribution from the eROSITA bubbles within $\sim 3 \times r_{200}$, and the bubbles are absent inside $r_{200}$. This inspired the surface brightness analysis of the cluster in the $1.0-2.3$\,keV band, described in Sect.~\ref{sec:SB}.

\begin{figure}
	\centering
 \includegraphics[width=0.95\linewidth]{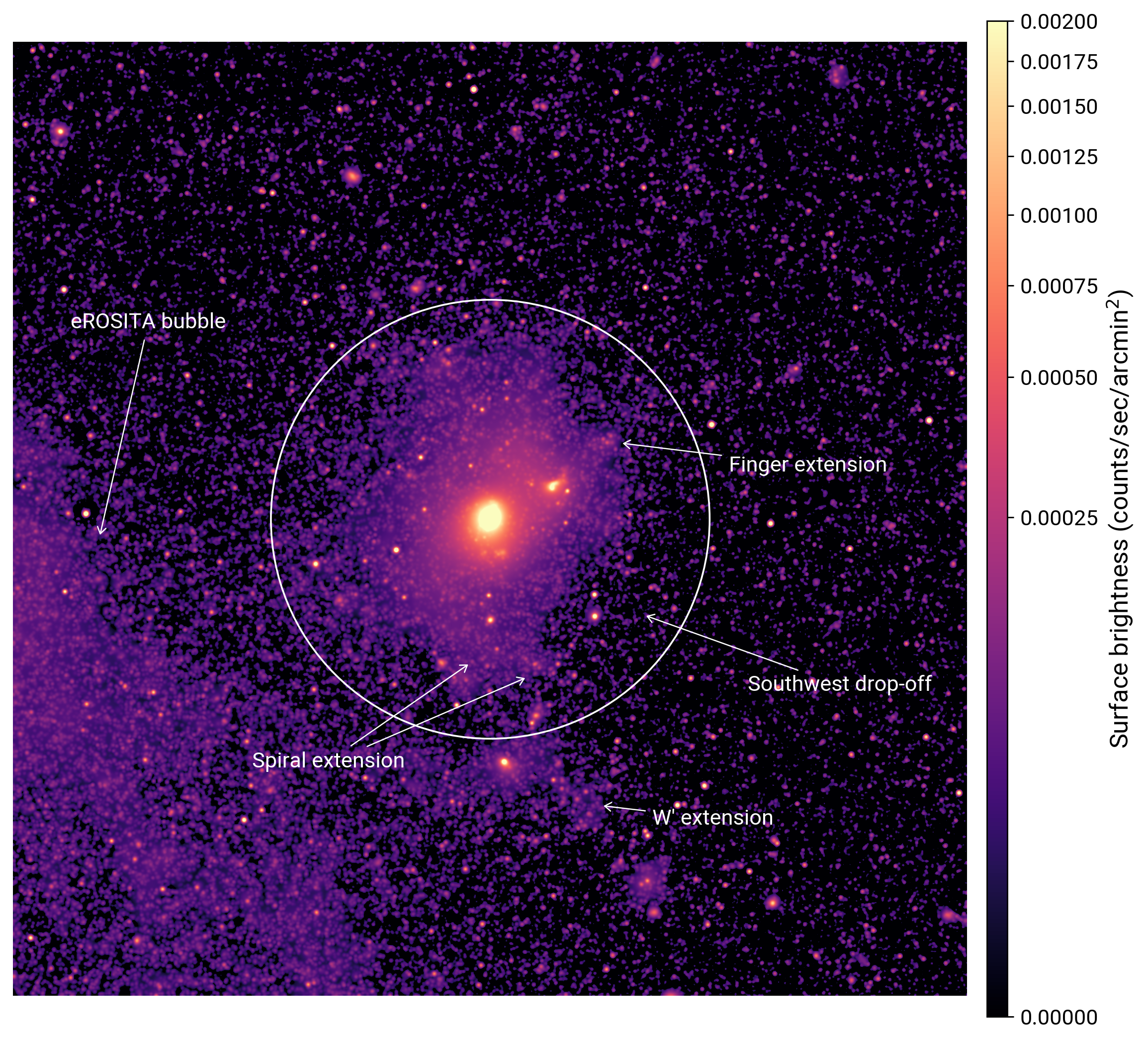}
	\caption{
 Wavelet filtered eRASS:5 image of the Virgo Cluster in the $0.3-2.0$\,keV energy band. The $r_{200}$ is shown in white. Features mentioned in the text are labeled.}
	\label{fig:wv_filt}
\end{figure}

As the focus of this study is the ICM emission, wavelet filtering was used for point source detection and enhancement of diffuse emission. The steps presented in \citet{Pacaud_2006} and \citet{Reiprich_2021} were followed to apply the wavelet filtering algorithm. Different parameters were set to manipulate the scales and thresholds depending on whether the goal of the filtering run was to detect point sources or emphasize significant ICM emission. For point source detection, the signal-to-noise threshold was lower than for ICM detection, a smaller background cell size was used, the maximum scale was lowered, and fewer iterations were required. The $0.3-2.0$\,keV wavelet filtered image can be seen in Fig.~\ref{fig:wv_filt}, while the ``blue band" $1.0-2.3$\,keV wavelet filtered image is in Fig.~\ref{fig:blue_img}. Point source detection was performed with the source detection tool \texttt{SExtractor} \citep{Bertin_1996}, with some modifications made after manual inspection. Additionally, background clusters were identified via the MCXC catalog \citep{Piffaretti_2011}. Our approach to point source detection also identifies small extended sources if they are above the signal-to-noise threshold of four; therefore, we have also excised additional background clusters that are not included in the MCXC catalog with this approach. Point sources and background clusters were then masked, with the radius identified via \texttt{SExtractor} and out to $r_{500}$, respectively, for all subsequent imaging manipulation and surface brightness analysis steps. 

\begin{figure}
	\centering
	\includegraphics[width=0.95\linewidth]{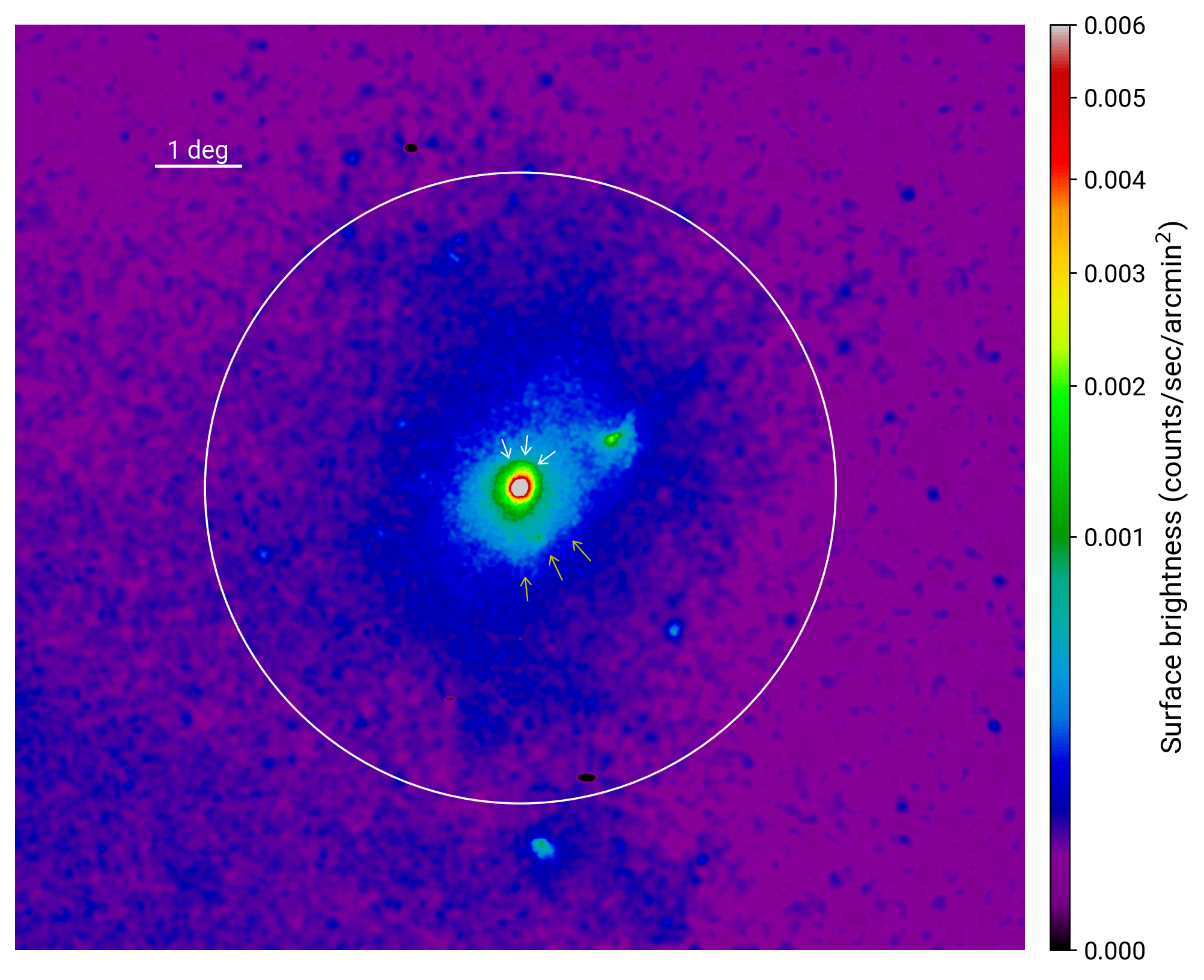}
	\caption{Point source excised, adaptively smoothed eRASS:5 image of the Virgo Cluster in the $0.3-2.0$\,keV energy band. The $r_{200}$ is shown in black. The cold fronts are marked with arrows, white for the northern and yellow for the southern cold front.}
	\label{fig:smoothed}
\end{figure}

Adaptive smoothing produces an image where the signal-to-noise ratio (S/N) at each pixel is approximately equivalent, meaning that fainter areas become more smoothed than brighter areas in an image. In this work, the task \texttt{asmooth} from the XMM-Newton Science Analysis System (\texttt{xmmsas}, \citealt{xmm}) with parameter \texttt{smoothstyle=`adaptive'} was used to apply adaptive smoothing to the PIB-subtracted, absorption- and exposure-corrected image with point sources masked. An S/N of ten was chosen, along with minimum and maximum values of zero and $25$ respectively for the normalized Gaussian convolvers. Masked point source areas were then effectively filled in with the signal level that surrounds them.  The point source excised, adaptively smoothed image is shown in Fig.~\ref{fig:smoothed}.

\begin{figure}
	\centering
	\includegraphics[width=0.95\linewidth]{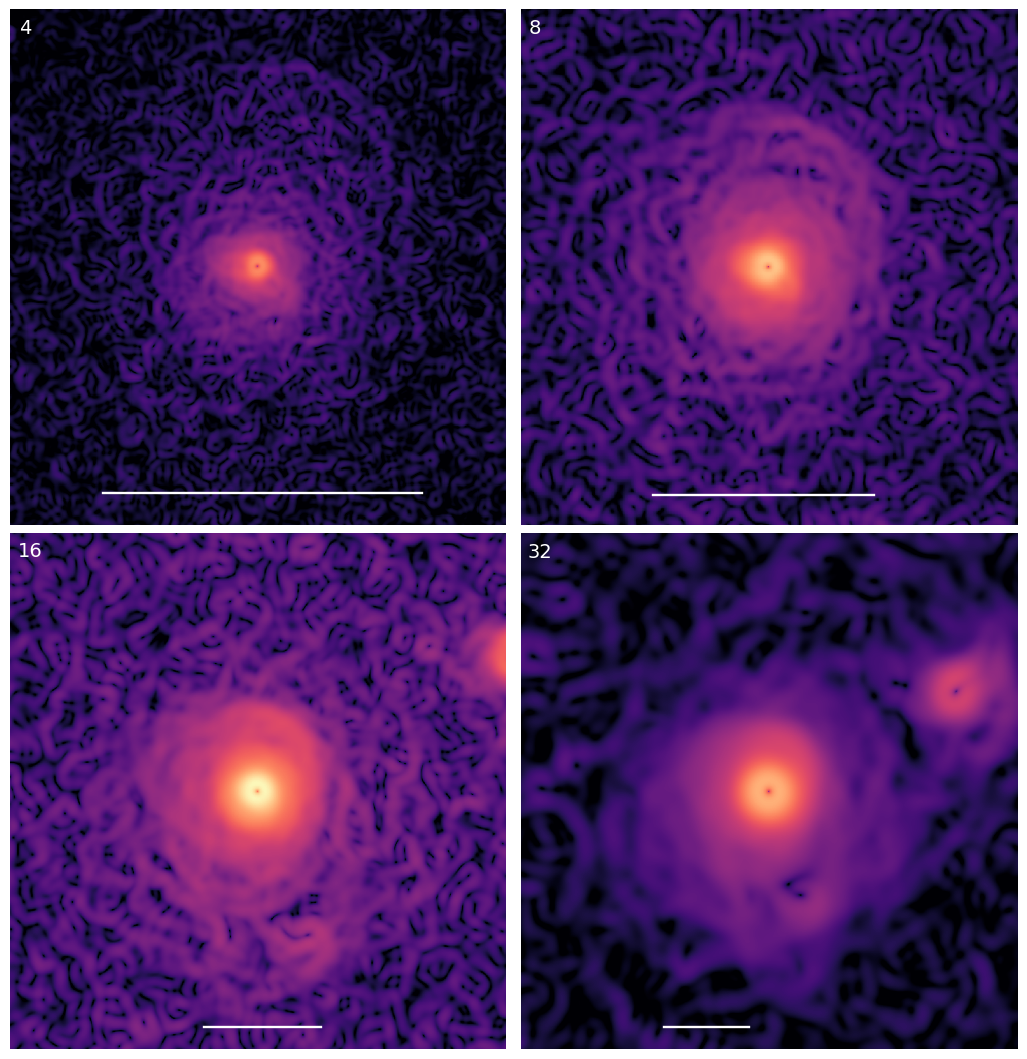}
	\caption{GGM filtered images with $\sigma = 4, 8, 16,$ and $32$\, pixels. The line has a length of 30 arcminutes, or 139.5\,kpc.}
	\label{fig:GGM}
\end{figure}

Gradient filtering has previously been used to examine features in clusters and simulations \citep{Sanders_2016, Walker_2016}. The Gaussian gradient magnitude (GGM) filter was applied following \citet{Sanders_2016} using \texttt{scipy} \citep{scipy} and $\sigma=1, 2, 4, 8, 16, 32$ detector pixels on an image with cosmetically removed point sources. A sample of the images that best show edges is in Fig.~\ref{fig:GGM}. The $\sigma = 4$ map shows the well-known M87 arms, while the $\sigma = 8$ map reveals a spiral edge feature to the northwest. On the largest scales, an extension to the southwest is prominent.

Looking at the wavelet filtered and adaptively smoothed images in Figs.\,\ref{fig:wv_filt} and \ref{fig:smoothed}, faint X-ray emission is visible out to $\sim 4-5^\circ$ from M87, beyond the virial radius, with the most conspicuous emission to the south out to M49 and beyond. An exception to this is the southwestern side, where there is a noticeable emission drop-off prior to the virial radius, a feature which is especially clear in the wavelet filtered image. To the south within the virial radius, the ICM curves in an extension to the southwest in a spiral shape. Further south, the emission halo around M49 appears more extended to its southern edge than to its northern, and is joined by an apparent extension, or tail, which curves off of the halo to the southwest in the direction of NGC 4261. To the north of the cluster, the emission appears more irregular than in the other directions. The emission seems to clump into a number of extensions stretching away from the cluster, with noticeable examples to the northeast and in the radial direction beyond M86. This finger-like tail protruding from M86, which appears out to a distance of 80\,arcminutes or 370\,kpc from the center of the galaxy, has been observed previously with other instruments. Using Chandra data, \citet{Randall_2008} asserted that the tail is related to ram pressure stripping as M86 falls into the Virgo Cluster and interacts with the ICM. To the southeast, the eROSITA bubble emission dominates all images, contaminating the ICM emission of Virgo. Overall, the cluster structure does not appear spherical or symmetrical.

\subsection{X-ray surface brightness analysis} \label{sec:SB}

To acquire the surface brightness profiles, \texttt{PyProffit} \citep{Eckert_2020} was used. For the full azimuthal profile, and subsequent profiles divided into eighths, 656 concentric annuli centered at the X-ray surface brightness peak, $(187.705^\circ, +12.391^\circ)$, and stretching to an outermost radius of 700\,arcminutes or $\sim 3r_{200}$ were utilized. Their radial bin distribution is logarithmic and has a minimum bin size of 26 arcseconds, approximately the same as the survey average point spread function (PSF) of eROSITA \citep{Predehl_2021}.

To estimate the magnitude of the cosmic X-ray background (CXB), two box regions of dimension $1.8^\circ$ by $3.1^\circ$ were extracted from outside $3r_{200}$ to the northwest to avoid cluster emission and contribution from the northern eROSITA bubble. The average value for 0.3-2.0\,keV was found to be $2.8 \times 10^{-3}$\,cts/s/arcmin$^2$, and for $1.0-2.3$\,keV, $1.4 \times 10^{-3}$\,cts/s/arcmin$^2$.

A single $\beta$-model \citep{Cavaliere_1976, Voit_2005}:

\begin{align}\label{eq:beta}
	S_\mathrm{X}(R) = S_\mathrm{X} (0) \left(1 + \frac{R^2}{r_\mathrm{c}^2}\right)^{-3\beta + \frac{1}{2}} + K ,
\end{align}
was fit to the $0.3-2.0$\,keV azimuthal profile out to 300 arcminutes, where $R$ is the projected distance from cluster center, $S_\mathrm{X} (0)$ is the normalization factor, $r_c$ is the core radius, $\beta$ determines the slope of the surface brightness profile, and $K$ is the background value. The best-fit $r_c$ and $\beta$ can be found in \autoref{tab:sb_ns_discs}. This simple model was chosen to allow for comparison with previous ROSAT publications, which also made use of the single $\beta$-model and comprehensive Virgo data. The best-fit slope of $0.43 \pm 0.02$ is in good agreement with ROSAT \citep{Nulsen_1995}, which cites $\beta = 0.46$, but the core radius of $47 \pm 5$\,arcseconds is about half the ROSAT value of 115 arcseconds. This difference seems mainly due to the poor fit of the single $\beta$-model to the innermost $\sim 2$\,arcminutes of the cluster combined with eROSITA's much better PSF as compared to ROSAT; additional factors could be different treatment of point sources, background clusters, and the central AGN between this work and \citet{Nulsen_1995}.

Although the spherical $\beta$-model allows us to directly compare to ROSAT results, we can also achieve a better fit to cluster emission by allowing for ellipticity in the model. This has the same form except that $R$ becomes
\begin{equation}
    R = \frac{(x_e^2 (1 - \epsilon)^2 + y_e^2)^{1/2}}{1 - \epsilon},
\end{equation}
where $\epsilon$ is the eccentricity and the coordinates are defined as 
$x_e = (x - x_0) \cos(\theta) + (y - y_0) \sin(\theta) $ and 
$ y_e = (y - y_0) \cos(\theta) - (x - x_0) \sin(\theta) $,
where $\theta$ is the ellipse's angle of rotation, and $x_0$ and $y_0$ are the central coordinates.

\begin{figure*}
	\centering
	\includegraphics[width=0.95\linewidth]{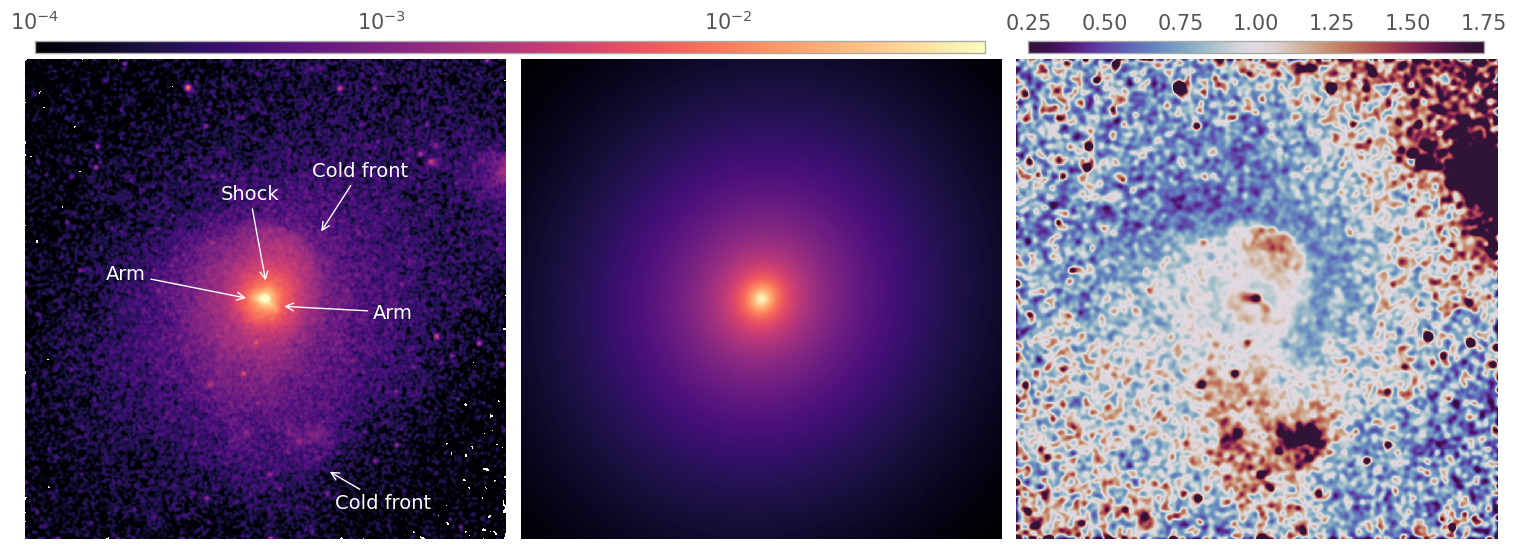}
	\caption{Series of eROSITA 0.3-2.0 keV images centered on M87 with half-size 65 arcminutes. Left: Surface brightness image smoothed with a sigma of 2 pixels, or 12 arcseconds. Point sources are present, but were removed for the fitting of the model. Middle: Best-fit $\beta$-model. Right: Flattened image, meaning the image divided by the model, smoothed with a sigma of 5 pixels, or 31 arcseconds.}
	\label{fig:flattened}
\end{figure*}

The best-fit elliptical model out to $r_{200}$ (also in \autoref{tab:sb_ns_discs}) was used to create a flattened image of the cluster center, meaning the surface brightness image was divided by the 2D model image. These data products can be seen in Fig.~\ref{fig:flattened}. In the flattened image, M87's characteristic arms are visible, as are several surface brightness edges. One is a weak shock identified in \citet{Forman_2007} with \textit{Chandra}. Two have been identified as candidate cold fronts in previous works and are explored further in the following section.

\subsubsection{Eight sectors}

\begin{figure}
	\centering
	\includegraphics[width=0.95\linewidth]{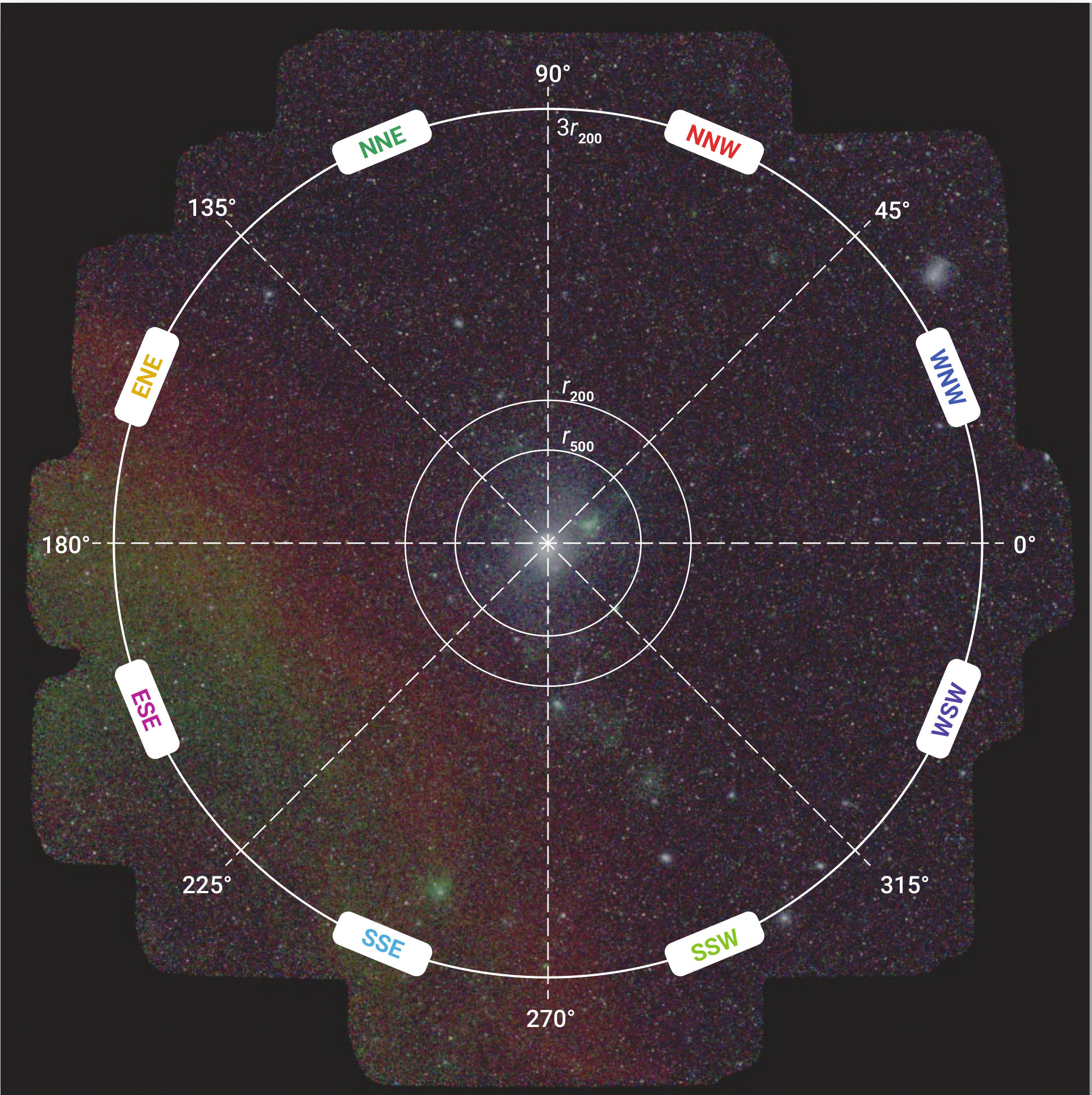}
	\caption{Image of the Virgo Cluster overlaid with the eight sector divisions and their corresponding names and colors. The annuli used for surface brightness profile extraction are not pictured. A number of radii are also shown and labeled.}
	\label{fig:SB_8profs_key}
\end{figure}

\begin{figure*}
	\centering
	\includegraphics[width=0.92\linewidth]{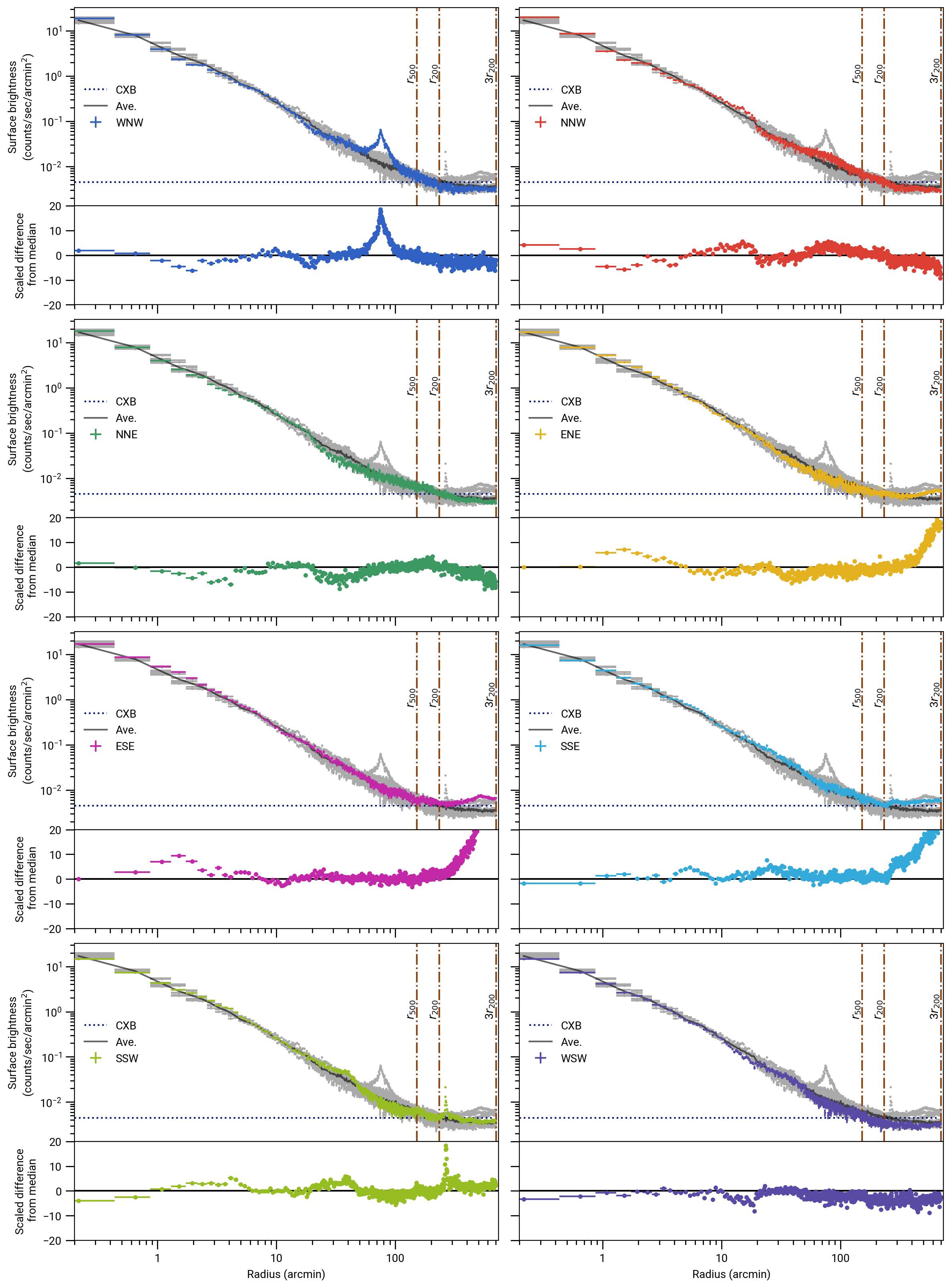}
	\caption{Eight surface brightness profiles in the $0.3-2.0$\,keV energy band, one for each sector identified in Fig.~\ref{fig:SB_8profs_key}. The other seven sectors for any given profile are plotted in gray for comparison, with the azimuthal average shown in black. The CXB measured to the northwest beyond $3r_{200}$ is shown in navy blue. A selection of characteristic radii are shown as dash-dotted vertical brown lines. The scaled difference to the median is the difference between each sector and the median divided by the uncertainty (``in sigmas'').}
	\label{fig:SB_8profs}
\end{figure*}

To better study features in any given direction, eight $0.3-2.0$\,keV surface brightness profiles were constructed. The choice of eight sectors was motivated by the decision not to bisect M86 or M49, the two largest subgroups, and to achieve a balance between information in more directions vs. to larger radii. Unlike, for example, ten sectors, eight sectors allows significant measurements to the outskirts of Virgo (beyond $r_{500}$) while offering more directional information than, for example, six sectors. The regions are shown in Fig.~\ref{fig:SB_8profs_key} and named for the eight half-wind compass points (north-northwest, east-northeast, etc.). The resulting surface brightness profile for each sector is seen in Fig.~\ref{fig:SB_8profs}. In this figure, each plot has a single profile for every sector, which is plotted in the color it corresponds to in Fig.~\ref{fig:SB_8profs_key}. The other seven sectors for any given profile are plotted in gray and the azimuthal average in black to offer a better idea of how the profiles compare. Profiles for the eight sectors were also created in the blue band of the RGB image, $1.0-2.3$\,keV, which is approximately free from eROSITA bubble contribution. These can be seen in Fig.~\ref{fig:SB_8profs_blue}, and are referred to as the ``blue'' profiles below.

The WNW sector contains a bump in the profile at $10$\,arcminutes corresponding to a surface brightness excess, and is followed by a drop-off at $20$\,arcminutes. A prominent peak in the profile is visible at $\sim 70$\,arcminutes or $330$\,kpc, corresponding to the X-ray halo around M86. The profile then gradually flattens out into the outskirts past $r_{200}$, where it shows a low level of surface brightness as compared to the other sectors. In the blue band, M86 is still prominent. 

In the NNW sector, which does not contain any of the brightest Virgo galaxies or groups, the profile displays a number of interesting features. Between $10$ and $20$\,arcminutes, there is an excess of brightness followed by a discontinuity in the profile. This behavior is often attributed to shocks or cold fronts, but to discern between the two, temperature and density profiles of the region are necessary \citep{Markevitch2007}. This particular feature was previously studied with \xmm, \textit{Chandra}, and Suzaku  data \citep{Urban2011,Werner_2016, Simionescu_2017} and identified as a cold front.
Another excess as compared to the other sectors is visible between $50$ and $100$\,arcminutes. This could be associated with the apparent extension of ICM emission in that direction. Past $r_{200}$, this sector shows among the lowest overall brightness when compared to the other sectors. This is expected because NNW is the furthest from the eROSITA bubbles. In the blue band, all of the same features are visible. 

In the NNE, a similar discontinuity as in NNW is visible at the same distance, $\sim 20$\,arcminutes, indicating that the cold front extends along the entire northern direction. There is a bump in the profile just past $r_{500}$ that could be due to the finger-like extension of the ICM observed in that direction. The outskirts past $r_{200}$ are at a similarly low level as in NNW with a similar shape in the full and blue profiles, indicating that there may not be substantial contribution by the eROSITA bubble in this region.

To the ENE, contribution from the eROSITA bubble becomes apparent both in the images and in the surface brightness profile, where the brightness increases from $r_{200}$ to $3r_{200}$.  The blue profile instead flattens out in this range.

The ESE sector shows a slow increase from $r_{200}$ to $3r_{200}$, a segment in which it registers the highest surface brightness of all sectors due to the eROSITA bubble. The blue profile for this region is not completely flat, but instead also begins to increase around 400\,arcminutes, implying some contribution from the eROSITA bubble in the $1.0-2.3$\,keV band to the ESE. However, the blue surface brightness in this range is still a factor three below the $0.3-2.0$\,keV surface brightness when normalized to the same central value, implying that most of the contribution from the eROSITA bubble is in the $0.3-1.0$\,keV band.

The SSE sector is brighter on average than other sectors around $40 - 50$\,arcminutes, and then drops off gradually. The eROSITA bubble clearly also affects the outer regions of this sector, which increases between $r_{200}$ and $3r_{200}$ and shows a prominent bump just after $r_{200}$ which becomes even more prominent in the blue band. This bump is prominent even following the removal of extended sources, or clumps, before production of the surface brightness profiles.

The SSW sector contains M49 and its extended ICM tail, features that appear in the profile in the form of a bump between $200$ and $300$\,arcminutes. A discontinuity, again possibly indicative of either a shock or cold front \citep{Markevitch2007}, is present in this profile at the same location as in the SSE around $40 - 50$\,arcminutes, except that it appears more pronounced in this sector.

Finally, to the WSW, the profile records the lowest values of all segments between $10-20$\,arcminutes, indicating that the cluster emission is not stretched along this direction, which matches our imaging analysis. There is a minor bump and drop around $50$\,arcminutes, perhaps indicative of the same feature present in the SSW and SSE. 

Overall, if we compare the blue and full surface brightness profiles normalized to the same central values, all of the same features observed in the full profile are present in the blue band. This confirms that these features are not due to the eROSITA bubble contribution.

\subsubsection{Cold fronts}

Two surface brightness discontinuities are apparent in the eight sector analysis, including one to the north that was previously identified in full by other X-ray instruments, and another to the south that is visible in full for the first time in this work. Both are marked in Figure~\ref{fig:smoothed}. We extracted surface brightness profiles across these discontinuities (Fig.~\ref{fig:northern_sector} and Fig.~\ref{fig:southern_sector}) in order to better investigate their properties.

At the location of the discontinuity found with eROSITA in the NNE and NNW sectors, \xmm, Suzaku, and \textit{Chandra} all record the same feature at this distance from cluster center to the north. These other instruments did not all cover the same portion of this feature; \citet{Simionescu_2010} utilizes an \xmm mosaic of the inner 200\,kpc, \citet{Urban2011} studies the cluster in a strip of pointings directly north from M87 with \xmm, \citet{Simionescu_2017} uses a strip of Suzaku pointings to the north, and \citet{Werner_2016} uses a northwestern \textit{Chandra} pointing to chart the sharpest part of the front.
The location of the discontinuity, at $19.1$\,arcminutes or 90\,kpc from cluster center, is identical across instruments. Although spectral analysis is outside the scope of this work, \citet{Simionescu_2010} uses spectral analysis to derive the temperature, density, and pressure profiles for this region, finding that the discontinuity is evidence of a cold front rather than a shock. The signature of cold fronts is a discontinuous drop in temperature paired with a jump in density, while shocks show an increase in temperature, density, and pressure \citep{Zinger_2018}. The analysis by \citet{Werner_2016} finds that it is sharper to the north than the west, which is in agreement with our surface brightness findings.

\citet{Simionescu_2010} also shows that the cold front is not centered spherically around M87, a conclusion supported by using the GGM and flattened images and experimentation with the eROSITA surface brightness of different regions around the location of the cold front to best determine its location. The cold front location is marked by a series of white arrows in Fig.~\ref{fig:smoothed}. The front is closer to M87 in the west than the east, matching the spiral morphology of gas sloshing in simulations \citep{Markevitch2007}. A single $\beta$-model with a discontinuity was fit to the eROSITA data in order to quantify the cold front. This model is found by scaling a standard single $\beta$-model by a sigmoid function. This takes the form

\begin{align}\label{eq:beta_disc}
	\left(1 - \frac{a}{1.0 + \exp(b(R - R_0))}\right) \times S_\mathrm{X} (0) \left(1 + \frac{R^2}{r_\mathrm{c}^2}\right)^{-3\beta + \frac{1}{2}} ,
\end{align}
where $a$ is the relative decrease of the surface brightness at the discontinuity, $b$ is the steepness of the jump, and $R_0$ is the location of the discontinuity. The values for the best fit can be found in \autoref{tab:sb_ns_discs} and the surface brightness profile across the discontinuity and model are shown in Fig.~\ref{fig:northern_sector}. We measure a surface brightness decrease at the discontinuity of $41\%$.

\begin{table}
	\centering
        \caption{Best-fit parameters to surface brightness profiles. }
	\begin{tabular}{cllll}
  \hline
   & Total Sph. & Total Ell. & North CF & South CF \\
  \hline
   $r_\mathrm{c}$\tablefootmark{a} & 47 $\pm$ 5 & 94 & - & - \\
  $\beta$ & 0.43 $\pm$ 0.02 & 0.42 & - & - \\
  $\theta$ & - & 1.035 & - & -\\
  $\epsilon$ & - & 0.07 & - & -\\
  $a$ & - & - & 0.41 $\pm$ 0.01 & 0.24 $\pm$ 0.02 \\
  $b$ & - &- & 1.29 $\pm$ 0.14 & 0.68 $\pm$ 0.22 \\
  \hline
 \end{tabular}
 \tablefoot{The total is fit to both spherical and elliptical $\beta$-models, while the ``north CF'' and ``south CF'' profiles are across the cold fronts and fit to a $\beta$-model with a discontinuity.
 \tablefoottext{a}{The core radius $r_\mathrm{c}$ is in units of arcseconds.} }
	\label{tab:sb_ns_discs}
\end{table}

To the south, the feature between $40-60$\,arcminutes ($\sim$220\,kpc) from center, with the distance dependent on the sector in which it is found (SSW, SSE, WSW) looks similar to the cold front feature in the north, although less sharp. A surface brightness discontinuity within this range, presumed to be the same feature, is also present in Suzaku data. \citet{Simionescu_2017} performs spectral analysis of two discontinuities, one at 233\,kpc in the western arm and one at 280\,kpc to the south, and finds that their temperature profiles are also indicative of a cold front. These two discontinuities are along the single discontinuity that we find, which is not spherical with respect to M87 but is closer to it in the west than the south. Future work should confirm that the discontinuity is consistent with a cold front along its entire extent.

We fit Eq.~(\ref{eq:beta_disc}) to the southern profile and found a less steep and less drastic decrease than the northern cold front. Best-fit values can be seen in \autoref{tab:sb_ns_discs} with the profile shown in Fig.~\ref{fig:southern_sector}. The off-set cold fronts in both the northern and southern directions could together be evidence of large-scale gas sloshing in the Virgo Cluster, a phenomenon caused by the infall of subclusters to the cluster core that displace the cool gas \citep{Markevitch2007}. Similar evidence for this type of gas sloshing in the Virgo Cluster was found in \citet{Simionescu_2007, Simionescu_2010, Gatuzz_2022}.

\subsubsection{Gas clumping}
\begin{figure}
	\centering
	\includegraphics[width=0.85\linewidth]{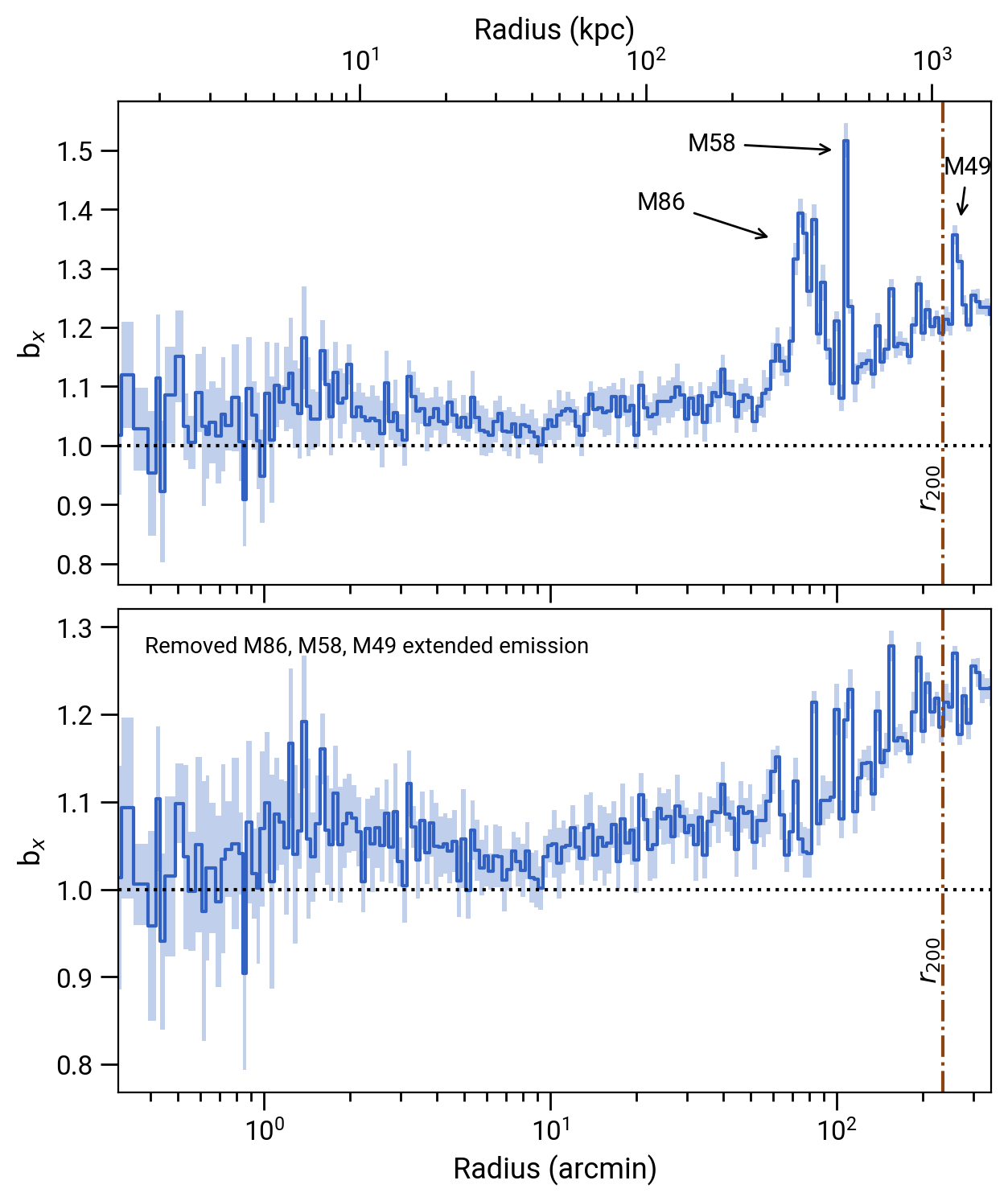}
	\caption{Emissivity bias, $b_X$, as a function of distance from cluster center. The top panel is the result after removing point sources and background clusters. The bottom panel is after additional removal of the extended gas halos around M86, M58, and M49 to show the upward trend still exists after removal of prominent sources.}
	\label{fig:emiss_bias}
\end{figure}

Based on \citet{Zhuravleva_2013}, which shows that the median surface brightness in simulated galaxy clusters is robust against gas inhomogeneities, \cite{Eckert_2015} proposed a method for recovering unbiased density profiles using the azimuthal median of the surface brightness in radial annuli. They apply the method to 31 clusters observed with ROSAT/PSPC. In this work, we applied the method, which involves making use of a 2D binning algorithm based on Voronoi tessellation \citep{Cappellari_2003}, to eROSITA data. The \texttt{VorBin} python package was used (see Sect.\, 3.3 of \cite{Eckert_2015} for further detail on methods). We computed the median surface brightness profile from the binned image and compared it with the surface brightness profile obtained by averaging. The comparison of the two methods can be seen in Fig.~\ref{fig:median_sb}. From these profiles, we computed the emissivity bias,
\begin{equation}
    b_X = \frac{S_{\rm X, mean}}{S_{\rm X, median}} ,
\end{equation}
a measure used in \cite{Eckert_2015} as a projected 2D surface brightness distribution proxy for the true 3D clumping factor of the gas. The emissivity bias obtained using the median and mean surface brightness profiles is plotted as a function of radius in Fig.~\ref{fig:emiss_bias}, out to 1$\sigma$ detection significance. Our results match the trend found in \cite{Eckert_2015}; the emissivity bias is low ($\sim 1.1$) in the inner regions of the cluster, and increases with radius in the outskirts ($\sim 1.2$). \citet{Mirakhor_2021} studies the clumping in Virgo with \textit{XMM-Newton} in the northern direction, and also finds mild clumping with a slight increasing trend. They report that their point source detection algorithm was likely able to exclude a significant fraction of clumps; this could well be the case in this work. As in that study, we also find that the exclusion of extended sources (primarily M86, M49, and M58) decreases the clumping factor significantly. After the exclusion of the emission around these three Virgo member galaxies, the measurement is consistent with the findings of \citet{Mirakhor_2021}. When the emissivity bias is calculated in the $1.0-2.3$\,keV band, the same trend exists, with the bias increasing to $\sim 1.2$ past $r_{200}$, implying that the trend is due to true cluster emission rather than foreground contribution from the eROSITA bubble.

Cosmological simulations predict that gas streams along cosmic web filaments can penetrate the inner regions of clusters, with gas clumping levels higher along filamentary directions \citep{Zinger_2016, Vazza_2013}. \citet{Simionescu_2017}, using the strip of Suzaku pointings, finds that gas density is higher along the north-south axis of Virgo than the east-west. They also find a steep increase in gas mass fraction beyond the virial radius, especially along the north-south axis. These findings support the idea that a cosmic web filament, complete with gas clumping, may run north-south in the Virgo Cluster, an idea also reinforced by the distribution of galaxies in the local Universe \citep{Castignani_2022}. \citet{Mirakhor_2021}, on the other hand, finds uniform clumping factor measurements in all directions when considering all available \xmm pointings.

To investigate whether eROSITA data indicates significant differences in clumping in varying directions, we also calculated the emissivity bias as a function of radius within the eight sectors identified in Fig.~\ref{fig:SB_8profs_key}. Larger radial bin widths and fewer bins than for the azimuthal assessment were used so as to achieve significant measurements out to $\sim r_{200}$ in every direction. For this calculation, the extended halos around M86, M49, and M58 were removed.

\begin{figure}
    \centering
    \includegraphics[width=0.85\linewidth]{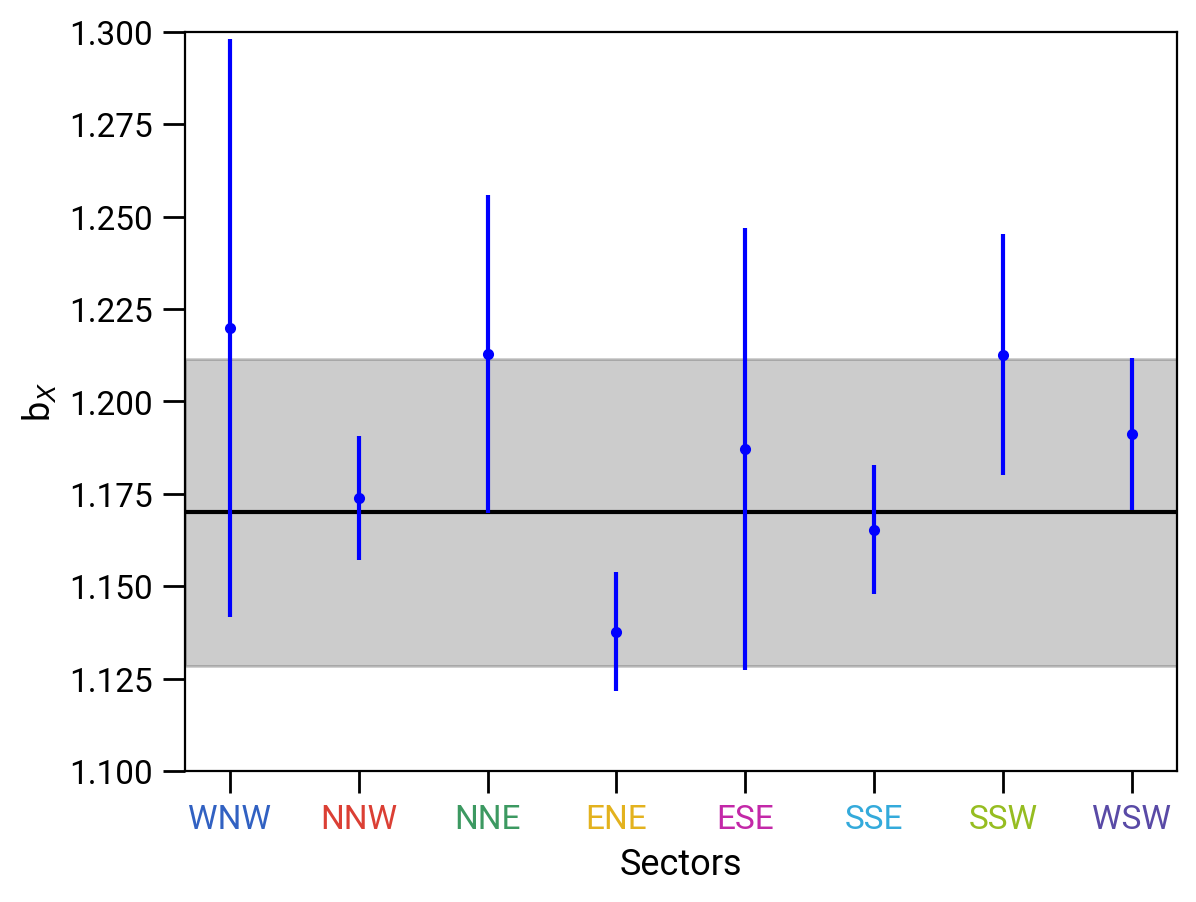}
    \caption{Azimuthal variation in the gas clumping in the radial range 0.75$r_{500}$ to 1.2$r_{200}$ for all eight sectors. The black line represents the average value. The clumping factor measurements are consistent, with no apparent north-south enhancement.}
    \label{fig:emiss_bias_all}
\end{figure}

The emissivity bias of each individual sector, shown in corresponding color in Fig.~\ref{fig:emiss_bias8} with the median in black, follows the same trend as the total profile. The values fluctuate around 1.0-1.1 until $\sim 60$\,arcminutes, after which point they increase to a median value of 1.2 just beyond $r_{200}$. Half of the sectors have single bins that spike to 1.3-1.4; namely, ENE, WSW, ESE, and WNW, all of which span the east-west axis of the cluster. However, no one sector is systematically higher than another. When the median of the four north-south axis sectors and four east-west axis sectors is compared (see Fig.~\ref{fig:emiss_bias_NS}), the measurements are within uncertainties of one another. That is, there is no apparent excess of clumping in the north-south axis as compared to the east-west out to $r_{200}$. Fig.~\ref{fig:emiss_bias_all} shows the good agreement among $b_X$ measurements in each sector beyond 0.75$r_{500}$.

Overall, the effect of clumping in the Virgo Cluster appears to be mild and broadly consistent in each radial direction. Moreover, with the full and, at the same time, detailed eROSITA view, we are not only able to measure the bias out to $\sim$$r_{200}$, but are also able to trace back the causes for stronger bias to individual sources. That is, the largest values of the emissivity bias occur at the locations of prominent galaxies and groups within Virgo, such as M86 at $\sim 100'$, M58 just beyond it, and M49 at $\sim 300'$. 

\section{Deprojected gas density and mass}\label{sec:deprojection}
Using \texttt{PyProffit}, deprojection was performed in order to extract gas density and gas mass estimates. We follow \citet{Eckert_2020}, which gives details of the application of the deprojection to eROSITA simulations. To calculate the conversion factor between count rate and emissivity, we assume an abundance of $0.2 \,\mathrm{Z}_{\odot}$\footnotemark and a temperature of $2.3\,$keV based on \citet{Simionescu_2017}, and the median $N_\mathrm{H}$ across the analyzed area from Sect.\,\ref{sec:data}. The conversion factor from count rate to emission measure is highly insensitive to temperature and abundance in this case. The multiscale decomposition deprojection was then performed. This allows for the extraction of the gas density profile, and subsequent calculation of the gas mass profile, which is the deprojected gas density integrated over a given volume. 

\begin{figure}
	\centering
	\includegraphics[width=0.85\linewidth]{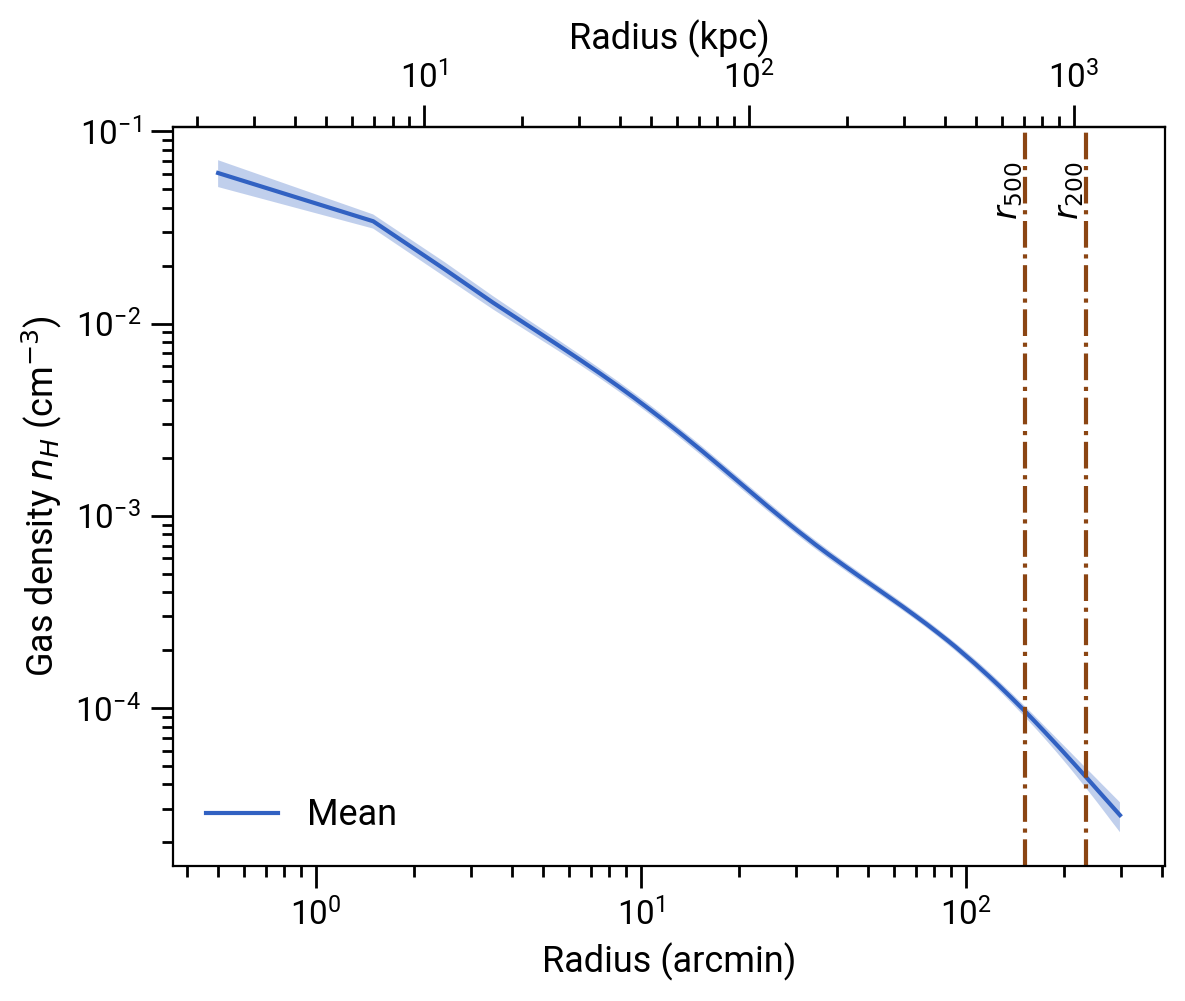}
	\caption{Deprojected gas density profile of the cluster out to 300 arcminutes. }
	\label{fig:gas_density}
\end{figure}

\begin{figure}
    \centering
    \includegraphics[width=0.85\linewidth]{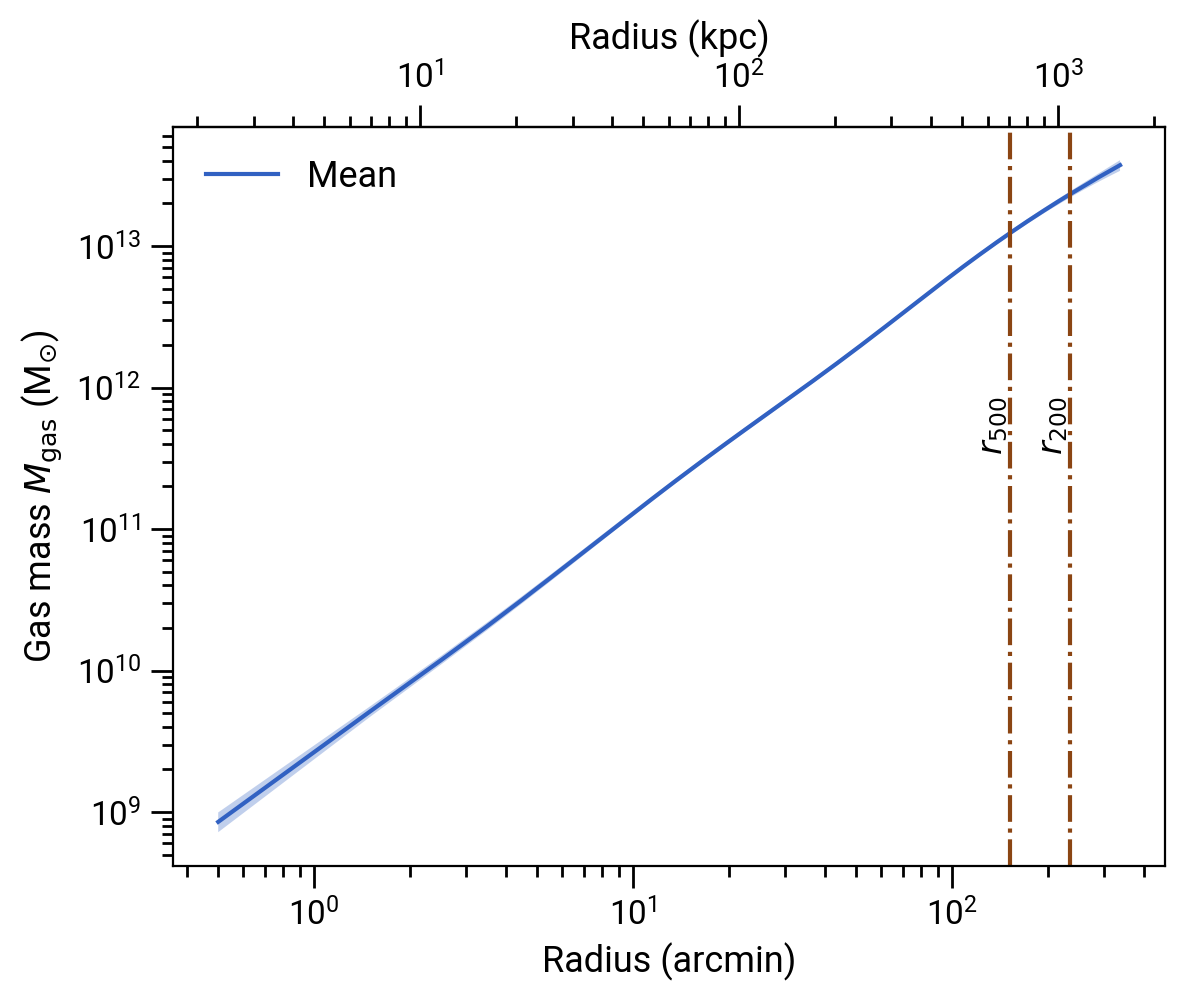}
	\caption{Deprojected gas mass profile of the cluster out to 300 arcminutes. }
	\label{fig:gas_mass}
\end{figure}

We applied these steps to the mean surface brightness profile. The gas density profile is plotted in Fig.~\ref{fig:gas_density}, and the gas mass in Fig.~\ref{fig:gas_mass}.

The hydrogen number density at 0.5\,arcminutes is estimated to be $0.06 \pm 0.01$\,cm$^{-3}$ and the density at $r_{200}$ = 1.08\,Mpc is $(4.37 \pm 0.51) \times 10^{-5}$\,cm$^{-3}$. The density profile approximately follows a power-law model $n_H \propto r^{-k}$ where $k = 1.25 \pm 0.05$ between 1.5 and 100 arcminutes. The slope is flatter inside 1.5\,arcminutes. The overall shape of the density profile matches typical profiles measured by ROSAT in that most clusters have a steepening profile beyond $\sim r_{500}$ \citep{Eckert_2012}. 

The deprojected mean profile yields gas masses $M_{\mathrm{gas},r<r_{500}} = (1.05 \pm 0.32) \times 10^{13}$\,M$_\odot$ and $M_{\mathrm{gas},r<r_{200}} = (1.98 \pm 0.60) \times 10^{13}$\,M$_\odot$. The gas mass profile has a best-fit power-law model with slope $k = -1.75 \pm 0.04$. Scaling relations from \citet{Arnaud_2005} predict a total gravitating mass of $M_{200} = 1.4 \times 10^{14}$\,M$_\odot$ for a cluster temperature of $2.3$\,keV, while \citet{Simionescu_2017} calculates an estimated virial mass of $(1.05 \pm 0.02) \times 10^{14}$\,M$_\odot$. This places our $M_{\mathrm{gas},r<r_{200}}$ gas mass estimate at approximately $14-19\%$ of the predicted total mass. Typical estimates of gas fraction in clusters are on the same order as the cosmic baryon fraction, $\sim 15\%$ \citep{Planck_2020}, though \citet{Simionescu_2017} finds increasing gas fraction with radius in Virgo, greater than $20\%$ for some radial directions beyond $r_{500}$.  When the deprojection is performed only for portions of the cluster that are not affected by the eROSITA bubble (i.e., northern and western segments), estimated masses are $\sim 10\%$ lower, still within reported uncertainties of the mean profile. 

It should be noted that the deprojection process requires the assumption of spherical symmetry, which we have shown in the rest of this work is a poor assumption in the case of Virgo due to its asymmetrical structure outside of the inner region. It is promising, however, that the gas clumping in all directions is shown to be mild and consistent. The non-uniformity of clumpy gas leads to an overestimate of density and therefore of mass; if clumping is mild and uniform, as our analysis shows, the resulting density and mass is less biased. \citet{Eckert_2015} finds typical $b_X$ values for their cluster sample that broadly agreed with those found here out to $r_{200}$, although with larger values and uncertainties at higher radii. Their clumping factors resulted in a gas mass overestimate in that work of $\sim 10\%$ at $r_{200}$. Therefore, the impact in this case, with lower $b_X$, would be $\leq 10\%$ gas mass bias above the true value due to gas clumping.

\footnotetext{\citet{Simionescu_2017} uses the abundance table of \citet{Lodders_2003}, but this difference plays a negligible role in determining conversion factor.}

\section{M49 region}\label{sec:M49}
A bright extended region, that is, to our knowledge, first studied in X-ray here with eROSITA, is located to the southwest of the M49 group, bent at a $50^\circ$ angle below the R.A. axis. It spans a projected length of 320 kpc ($1.1^\circ$), is 140 kpc ($0.5^\circ$) wide, and is approximately 1.3\,Mpc from the center of the Virgo Cluster, outside $r_{200}$. A surface brightness profile across the extension shows the morphology as singly peaked (in contrast to double-peaked stripping tails, see e.g., \citealt{Ge_2023}). To determine the significance of this enhancement, the surface brightness and propagated error was extracted from a number of $20' \times 20'$ square regions along the extension (so-called source boxes) and just outside of it (so-called background boxes). A mean value for the source boxes and for the background boxes was calculated. The error was found with
\begin{equation}
    \textrm{Error} = \sqrt{\sigma_\mathrm{src}^2 + \sigma_\mathrm{bkg}^2} ,
\end{equation}
where $\sigma_\mathrm{src}$ is the standard error of the mean surface brightness of the source boxes and $\sigma_\mathrm{bkg}$ is the standard error of the mean surface brightness of the background boxes. The relative difference between the extension and background is found to be 41.75\%, or a significance of 3.15$\sigma$.

We searched the projected area around this feature for clues as to its origin. It is coincident with the so-called W'-cloud, made up of elliptical galaxy NGC 4365 and several member galaxies, which has been previously studied in the optical and infrared \citep{deVaucouleurs_1961, Binggeli_1987, Mei_2007}. This group is located at $\sim$ 23\,Mpc \citep{Mei_2007} and is proposed to be falling into Virgo from behind. NGC 4365 and fellow group member NGC 4342 are coincident and thus likely responsible for the peaks in X-ray emission within the extension. The extended X-ray emission could be the W' intra-group medium, extended due to its fall into the cluster. Nearby, to the northeast of the extension, the M49 group is likely falling into the cluster from the south \citep{Su_2019}. \cite{Su_2019} finds an M49 stripped tail of smaller dimensions (70 kpc long by 10 kpc wide), which they argue is evidence of the brightest group galaxy (BGG) moving relative to the M49 group gas. This smaller tail associated with the BGG is also visible with eROSITA; the longer extension matches the morphology and direction of the smaller one. The longer extension could be a stripped tail of intra-group gas, as opposed to the smaller stripped tail of intergalactic gas.

\begin{figure}
	\centering
	\includegraphics[width=0.45\linewidth]{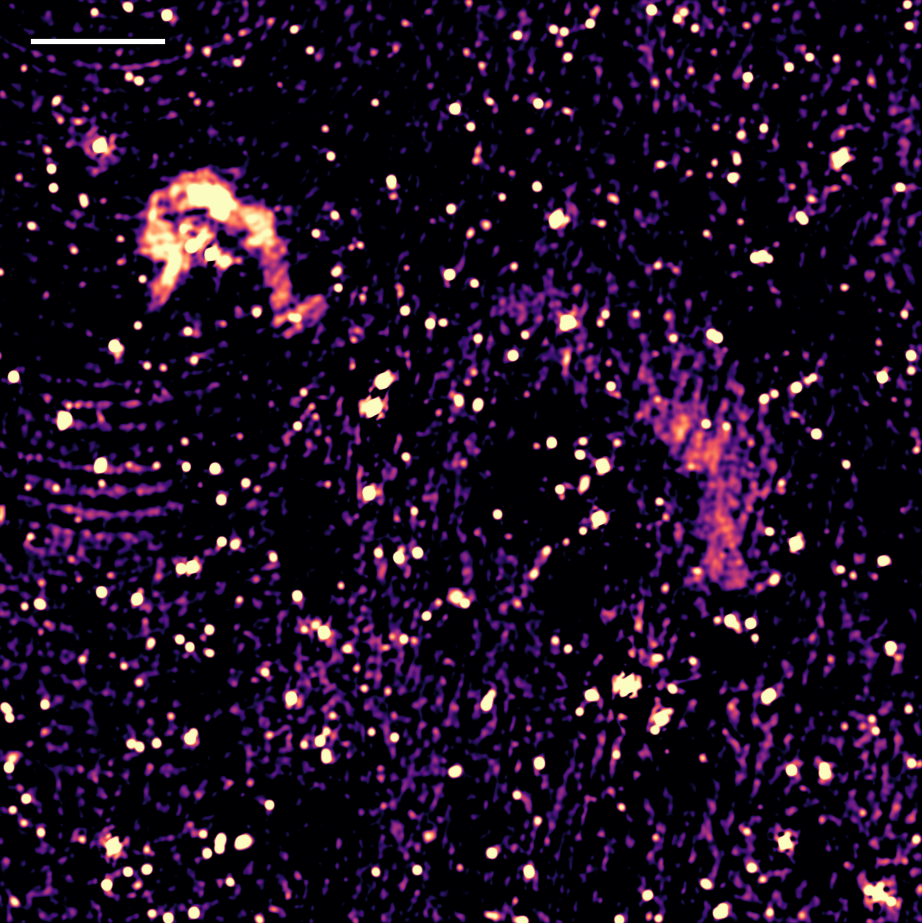}
        \includegraphics[width=0.46\linewidth]{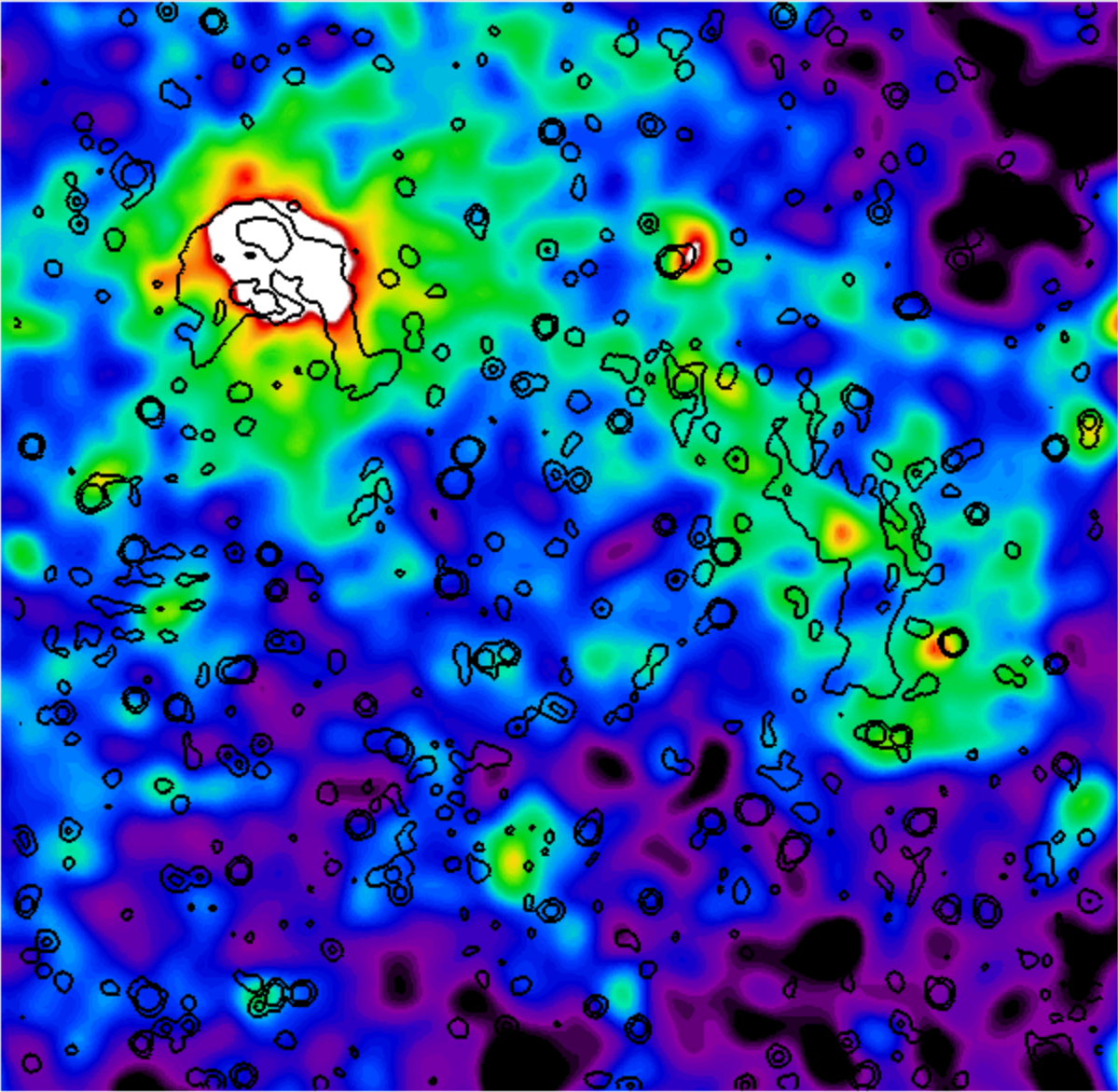}
        \caption{Image displaying the region to the southwest of M49 in radio with LOFAR (left) and in X-ray with 
        eROSITA (right). Contours of the radio image have been overlaid on the X-ray image for easier comparison, and images are displayed with the same frame. The white line in the upper left is 100\,kpc in length.}
	\label{fig:M49}
\end{figure}

A diffuse feature with a similar extent, location, morphology, and orientation as the X-ray feature was also recently identified in the radio band with LOw-Frequency ARray High-Band Antenna (LOFAR HBA) as part of ViCTORIA (Virgo Cluster multiTelescope Observations in Radio of Interacting galaxies and AGN) project \citep{Edler_2023}. The radio extension has a length of $\sim 1^{\circ}$, about 280\,kpc, and shares an orientation with the radio tails around M49, although no clear connection can be seen between structures. Figure~\ref{fig:M49} shows a cut-out of the region in radio, and in X-ray with radio contours overlaid, showing the similar structures of X-ray and radio emission. 

Based on its morphology and shared location, we suggest that the X-ray extension could be causally connected to the radio. The feature could be related to the in-falling of diffuse gas associated with the W' cloud group or stripped gas of the M49 group and its interaction with the ICM. The radio feature could possibly originate in an accretion shock or due to turbulence in the hot plasma \citep{vanWeeren_2019}. An alternative explanation for the radio feature, which does not, however, explain the similar morphology in the radio and X-rays, is that the radio emission is due to remnant AGN plasma related to a previous active phase of NGC\,4365 \citep{Spasic_2024}. A follow-up on its X-ray and radio spectral properties is required to gain a better understanding of this feature.

\section{Conclusion} \label{sec:conclude}

In this work, data collected during the four complete all-sky surveys and partial fifth survey of eROSITA is used to study the Virgo Cluster in an area spanning approximately 25$^\circ$ by 25$^\circ$. 
Data reduction and preparation include flare removal and corrections for PIB, variation of $N_{HI}$ across the sky area, and exposure differences between TM\,8 and TM\,9.

Different image manipulation techniques allow for the exploration of larger scale structure.
The RGB image reveals extended white-green emission around prominent galaxies and groups, as well as a gradient within the northern eROSITA bubble, which is mainly projected onto Virgo Cluster emission in the red and green bands (0.3-1.0\,keV). 
The bubble fades from blue-green to green ($0.6-1.0$\,keV) to red ($0.3-0.6$\,keV) as the distance grows from the Galactic center.
Wavelet filtering and adaptive smoothing reveal faint ICM emission in several directions out to $4-5^\circ$ from M87 with a larger radius to the north and south, and a notable drop-off in emission to the west prior to the virial radius.
The northern region of the Virgo Cluster contains a number of ICM extensions stretching away from the cluster, while a spiral-like extension can be seen to the south.

Surface brightness profiles were created for the full azimuth and eighths of the overall cluster in the $0.3-2.0\,$keV and $1.0-2.3\,$keV bands. The latter band choice arose from the apparent lack of contribution to the Virgo Cluster emission by the northern eROSITA bubble in the blue emission of the RGB image. 
To the north, in the NNW and NNE subsectors, a prominent discontinuity at $\sim 19$\,arcminutes marks a $41\%$ decrease in surface brightness.
This has been identified as a cold front in other studies that used spectral analysis.
To the south, most prominent in the SSW subsector and less sharp in SSE and WSW, another discontinuity in the profile is present at $\sim 48$\,arcminutes, though with just a $24\%$ decrease. 
This was also previously identified as a cold front at locations along it to the south and west; however, this is the first time the full extent of the front can be seen.
The off-set cold fronts to the north and south of the cluster support a picture of large-scale gas sloshing.
The $0.3-2.0$\,keV surface brightness profile to the east is dominated by the eROSITA bubble past $r_{200}$, while the $1.0-2.3$\,keV profile shows only a minor contribution from the bubble in this region. The surface brightness profiles to the west match the visual evidence of the images, displaying few notable features except for an excess at the location of the ICM surrounding M86.
All reported features (with the exception of the eROSITA bubble) are present and similarly significant in both the $0.3-2.0\,$keV and $1.0-2.3\,$keV bands. 

We make use of Voronoi tessellation to bin the data before extracting a median surface brightness profile. One would expect the median surface brightness profile to be robust against gas clumping down to the scale of the typical Voronoi cell size, unlike the mean surface brightness profile. The comparison of the two can be quantified by the emissivity bias, $b_X$, which is larger in the case of gas clumping. We find a low emissivity bias, about 1.1, in the inner regions, growing to about 1.2 in the outskirts of the Virgo Cluster. Peaks in the emissivity bias can be traced back to M86, M58, and M49, with an increasing trend still apparent following their removal. Emissivity bias was also calculated for each of eight sectors introduced for surface brightness analysis. Although half of the sectors show single bin peaks to values of 1.3-1.4, no single sector has consistently higher gas clumping. The emissivity biases in the outskirts in each direction, and consequently along the north-south and east-west axes, are consistent. The effect of gas clumping from this measure appears to be mild and radially uniform. This detailed measurement is the first of its kind for the full azimuth of Virgo, and offers a high resolution view of clumping that would be difficult to achieve in more distant clusters.

A deprojection analysis was performed in order to determine gas density and gas mass profiles. The result was a virial gas mass of $M_{\mathrm{gas},r<r_{200}} = (1.98 \pm 0.70) \times 10^{13}$\,M$_\odot$, which is on the order of 14-19\% of total mass estimates for Virgo, largely consistent with the expected gas mass fraction. This reported gas mass does not incorporate potential overestimates due to contribution from the eROSITA bubbles and due to gas clumping, which could each contribute a $\lesssim 10\%$ inflation.

 An X-ray extension south of M49, outside the virial radius of Virgo, has a length of $\sim 320$\,kpc.
The surface brightness of the extension has a relative difference to the background of $41.75\%$ and a lower-bound significance of $3.15 \sigma$. These values were determined using extracted surface brightness values along and just outside of the extension.
The X-ray emission is coincident with radio emission observed by LOFAR. The aligned morphology to the M49 BGG stripped tail, shared location with the W' cloud galaxies behind Virgo, and coincident radio emission could together be indication of an accretion process such as a shock or turbulence.

\begin{acknowledgements}
      This work is based on data from eROSITA, the soft X-ray instrument aboard SRG, a joint Russian-German science mission supported by the Russian Space Agency (Roskosmos), in the interests of the Russian Academy of Sciences represented by its Space Research Institute (IKI), and the Deutsches Zentrum für Luft- und Raumfahrt (DLR). The SRG spacecraft was built by Lavochkin Association (NPOL) and its subcontractors, and is operated by NPOL with support from the Max Planck Institute for Extraterrestrial Physics (MPE). The development and construction of the eROSITA X-ray instrument was led by MPE, with contributions from the Dr. Karl Remeis Observatory Bamberg \& ECAP (FAU Erlangen-Nuernberg), the University of Hamburg Observatory, the Leibniz Institute for Astrophysics Potsdam (AIP), and the Institute for Astronomy and Astrophysics of the University of Tübingen, with the support of DLR and the Max Planck Society. The Argelander Institute for Astronomy of the University of Bonn and the Ludwig Maximilians Universität Munich also participated in the science preparation for eROSITA. The eROSITA data shown here were processed using the eSASS software system developed by the German eROSITA consortium.

      HM thanks Dominique Eckert for his correspondence on the use of \texttt{PyProffit}.

     MB acknowledges support from the Deutsche Forschungsgemeinschaft under Germany's Excellence Strategy - EXC 2121 "Quantum Universe" - 390833306 and from the BMBF ErUM-Pro grant 05A2023. AV acknowledges funding by the Deutsche Forschungsgemeinschaft (DFG, German Research Foundation) -- 450861021. FdG acknowledges support from the ERC Consolidator Grant ULU 101086378.

\end{acknowledgements}

%
\bibliographystyle{aa} 
\bibliography{final_bib} 

\begin{thebibliography}{68}
\expandafter\ifx\csname natexlab\endcsname\relax\def\natexlab#1{#1}\fi

\bibitem[{Aghanim {et~al.}(2020)Aghanim, Akrami, Ashdown, Aumont, Baccigalupi,
  Ballardini, Banday, Barreiro, Bartolo, Basak, Battye, Benabed, Bernard,
  Bersanelli, Bielewicz, Bock, Bond, Borrill, Bouchet, Boulanger, Bucher,
  Burigana, Butler, Calabrese, Cardoso, Carron, Challinor, Chiang, Chluba,
  Colombo, Combet, Contreras, Crill, Cuttaia, de~Bernardis, de~Zotti,
  Delabrouille, Delouis, Di~Valentino, Diego, Doré, Douspis, Ducout, Dupac,
  Dusini, Efstathiou, Elsner, Enßlin, Eriksen, Fantaye, Farhang, Fergusson,
  Fernandez-Cobos, Finelli, Forastieri, Frailis, Fraisse, Franceschi, Frolov,
  Galeotta, Galli, Ganga, Génova-Santos, Gerbino, Ghosh, González-Nuevo,
  Górski, Gratton, Gruppuso, Gudmundsson, Hamann, Handley, Hansen, Herranz,
  Hildebrandt, Hivon, Huang, Jaffe, Jones, Karakci, Keihänen, Keskitalo,
  Kiiveri, Kim, Kisner, Knox, Krachmalnicoff, Kunz, Kurki-Suonio, Lagache,
  Lamarre, Lasenby, Lattanzi, Lawrence, Le~Jeune, Lemos, Lesgourgues, Levrier,
  Lewis, Liguori, Lilje, Lilley, Lindholm, López-Caniego, Lubin, Ma,
  Macías-Pérez, Maggio, Maino, Mandolesi, Mangilli, Marcos-Caballero, Maris,
  Martin, Martinelli, Martínez-González, Matarrese, Mauri, McEwen, Meinhold,
  Melchiorri, Mennella, Migliaccio, Millea, Mitra, Miville-Deschênes,
  Molinari, Montier, Morgante, Moss, Natoli, Nørgaard-Nielsen, Pagano,
  Paoletti, Partridge, Patanchon, Peiris, Perrotta, Pettorino, Piacentini,
  Polastri, Polenta, Puget, Rachen, Reinecke, Remazeilles, Renzi, Rocha,
  Rosset, Roudier, Rubiño-Martín, Ruiz-Granados, Salvati, Sandri, Savelainen,
  Scott, Shellard, Sirignano, Sirri, Spencer, Sunyaev, Suur-Uski, Tauber,
  Tavagnacco, Tenti, Toffolatti, Tomasi, Trombetti, Valenziano, Valiviita,
  Van~Tent, Vibert, Vielva, Villa, Vittorio, Wandelt, Wehus, White, White,
  Zacchei, \& Zonca}]{Planck_2020}
Aghanim, N., Akrami, Y., Ashdown, M., {et~al.} 2020, A\&A, 641, A6

\bibitem[{Ar{\'e}valo {et~al.}(2016)Ar{\'e}valo, Churazov, Zhuravleva, Forman,
  \& Jones}]{Arevalo_2016}
Ar{\'e}valo, P., Churazov, E., Zhuravleva, I., Forman, W.~R., \& Jones, C.
  2016, ApJ, 818, 14

\bibitem[{Arnaud(1996)}]{Arnaud_1996}
Arnaud, K. 1996, in Astronomical Society of the Pacific Conference Series, Vol.
  101, Astronomical Data Analysis Software and Systems V, 17

\bibitem[{{Arnaud} {et~al.}(2005){Arnaud}, {Pointecouteau}, \&
  {Pratt}}]{Arnaud_2005}
{Arnaud}, M., {Pointecouteau}, E., \& {Pratt}, G.~W. 2005, \aap, 441, 893

\bibitem[{{Asplund} {et~al.}(2009){Asplund}, {Grevesse}, {Sauval}, \&
  {Scott}}]{Asplund_2009}
{Asplund}, M., {Grevesse}, N., {Sauval}, A.~J., \& {Scott}, P. 2009, \araa, 47,
  481

\bibitem[{{Bertin} \& {Arnouts}(1996)}]{Bertin_1996}
{Bertin}, E. \& {Arnouts}, S. 1996, AAPS, 117, 393

\bibitem[{{Binggeli} {et~al.}(1987){Binggeli}, {Tammann}, \&
  {Sandage}}]{Binggeli_1987}
{Binggeli}, B., {Tammann}, G.~A., \& {Sandage}, A. 1987, \aj, 94, 251

\bibitem[{{B\"ohringer} {et~al.}(1994){B\"ohringer}, {Briel}, {Schwarz, R. A.},
  \& {et al.}}]{Bo1994}
{B\"ohringer}, H., {Briel}, U.~G., {Schwarz, R. A.}, \& {et al.} 1994, Nature

\bibitem[{{B\"ohringer} {et~al.}(1995){B\"ohringer}, {Nulsen}, {Braun}, \&
  {Fabian}}]{Bo_1995}
{B\"ohringer}, H., {Nulsen}, P.~E.~J., {Braun}, R., \& {Fabian}, A.~C. 1995,
  \mnras, 274, L67

\bibitem[{Brunner {et~al.}(2022)Brunner, Liu, Lamer, Georgakakis, Merloni,
  Brusa, Bulbul, Dennerl, Friedrich, Liu, Maitra, Nandra, Ramos-Ceja, Sanders,
  Stewart, Boller, Buchner, Clerc, Comparat, Dwelly, Eckert, Finoguenov,
  Freyberg, Ghirardini, Gueguen, Haberl, Kreykenbohm, Krumpe, Osterhage,
  Pacaud, Predehl, Reiprich, Robrade, Salvato, Santangelo, Schrabback, Schwope,
  \& Wilms}]{Brunner_2022}
Brunner, H., Liu, T., Lamer, G., {et~al.} 2022, Astronomy {\&} Astrophysics,
  661, A1

\bibitem[{{Bulbul} {et~al.}(2022){Bulbul}, {Liu}, {Pasini}, {Comparat},
  {Hoang}, {Klein}, {Ghirardini}, {Salvato}, {Merloni}, {Seppi}, {Wolf},
  {Anderson}, {Bahar}, {Brusa}, {Br{\"u}ggen}, {Buchner}, {Dwelly},
  {Ibarra-Medel}, {Ider Chitham}, {Liu}, {Nandra}, {Ramos-Ceja}, {Sanders}, \&
  {Shen}}]{Bulbul_2022}
{Bulbul}, E., {Liu}, A., {Pasini}, T., {et~al.} 2022, \aap, 661, A10

\bibitem[{{Cappellari} \& {Copin}(2003)}]{Cappellari_2003}
{Cappellari}, M. \& {Copin}, Y. 2003, MNRAS, 342, 345

\bibitem[{{Castignani} {et~al.}(2022){Castignani}, {Vulcani}, {Finn}, {Combes},
  {Jablonka}, {Rudnick}, {Zaritsky}, {Whalen}, {Conger}, {De Lucia}, {Desai},
  {Koopmann}, {Moustakas}, {Norman}, \& {Townsend}}]{Castignani_2022}
{Castignani}, G., {Vulcani}, B., {Finn}, R.~A., {et~al.} 2022, ApJS, 259, 43

\bibitem[{{Cavaliere} \& {Fusco-Femiano}(1976)}]{Cavaliere_1976}
{Cavaliere}, A. \& {Fusco-Femiano}, R. 1976, \aap, 49, 137

\bibitem[{{Churazov} {et~al.}(2001){Churazov}, {Br{\"u}ggen}, {Kaiser},
  {B{\"o}hringer}, \& {Forman}}]{Churazov_2001}
{Churazov}, E., {Br{\"u}ggen}, M., {Kaiser}, C.~R., {B{\"o}hringer}, H., \&
  {Forman}, W. 2001, \apj, 554, 261

\bibitem[{{de Vaucouleurs}(1961)}]{deVaucouleurs_1961}
{de Vaucouleurs}, G. 1961, \apjs, 6, 213

\bibitem[{Eckert {et~al.}(2020)Eckert, Finoguenov, Ghirardini, Grandis, Käfer,
  Sanders, \& Ramos-Ceja}]{Eckert_2020}
Eckert, D., Finoguenov, A., Ghirardini, V., {et~al.} 2020, The Open Journal of
  Astrophysics, 3

\bibitem[{{Eckert} {et~al.}(2015){Eckert}, {Roncarelli}, {Ettori}, {Molendi},
  {Vazza}, {Gastaldello}, \& {Rossetti}}]{Eckert_2015}
{Eckert}, D., {Roncarelli}, M., {Ettori}, S., {et~al.} 2015, \mnras, 447, 2198

\bibitem[{{Eckert, D.} {et~al.}(2012){Eckert, D.}, {Vazza, F.}, {Ettori, S.},
  {Molendi, S.}, {Nagai, D.}, {Lau, E. T.}, {Roncarelli, M.}, {Rossetti, M.},
  {Snowden, S. L.}, \& {Gastaldello, F.}}]{Eckert_2012}
{Eckert, D.}, {Vazza, F.}, {Ettori, S.}, {et~al.} 2012, A\&A, 541, A57

\bibitem[{Edler {et~al.}(2023)Edler, de~Gasperin, Shimwell, Hardcastle,
  Boselli, Heesen, McCall, Bomans, Brüggen, Bulbul, Chy{\.{z} }y, Ignesti,
  Merloni, Pacaud, Reiprich, Roberts, Rottgering, \& van Weeren}]{Edler_2023}
Edler, H.~W., de~Gasperin, F., Shimwell, T.~W., {et~al.} 2023, A \& A, 676, A24

\bibitem[{{Forman} {et~al.}(2007){Forman}, {Jones}, {Churazov}, {Markevitch},
  {Nulsen}, {Vikhlinin}, {Begelman}, {B{\"o}hringer}, {Eilek}, {Heinz},
  {Kraft}, {Owen}, \& {Pahre}}]{Forman_2007}
{Forman}, W., {Jones}, C., {Churazov}, E., {et~al.} 2007, \apj, 665, 1057

\bibitem[{Forman {et~al.}(2005)Forman, Nulsen, Heinz, Owen, Eilek, Vikhlinin,
  Markevitch, Kraft, Churazov, \& Jones}]{Forman_2005}
Forman, W., Nulsen, P., Heinz, S., {et~al.} 2005, ApJ, 635, 894

\bibitem[{{Freyberg} {et~al.}(2020){Freyberg}, {Perinati}, {Pacaud}, {Eraerds},
  {Churazov}, {Dennerl}, {Predehl}, {Merloni}, {Meidinger}, {Bulbul},
  {Friedrich}, {Gilfanov}, {Tenzer}, {Pommranz}, {Eckert}, {Schmitt}, {Brusa},
  \& {Santangelo}}]{Freyberg_2020}
{Freyberg}, M., {Perinati}, E., {Pacaud}, F., {et~al.} 2020, in Society of
  Photo-Optical Instrumentation Engineers (SPIE) Conference Series, Vol. 11444,
  Society of Photo-Optical Instrumentation Engineers (SPIE) Conference Series,
  114441O

\bibitem[{{Gatuzz} {et~al.}(2023){Gatuzz}, {Sanders}, {Dennerl}, {Liu},
  {Fabian}, {Pinto}, {Eckert}, {Russell}, {Tamura}, {Walker}, \&
  {ZuHone}}]{Gatuzz_2023}
{Gatuzz}, E., {Sanders}, J.~S., {Dennerl}, K., {et~al.} 2023, \mnras, 520, 4793

\bibitem[{{Gatuzz} {et~al.}(2022){Gatuzz}, {Sanders}, {Dennerl}, {Pinto},
  {Fabian}, {Tamura}, {Walker}, \& {ZuHone}}]{Gatuzz_2022}
{Gatuzz}, E., {Sanders}, J.~S., {Dennerl}, K., {et~al.} 2022, \mnras, 511, 4511

\bibitem[{Ge {et~al.}(2023)Ge, Sun, Nulsen, Sarazin, Markevitch, \&
  Schellenberger}]{Ge_2023}
Ge, C., Sun, M., Nulsen, P. E.~J., {et~al.} 2023, MNRAS, 525, 1365

\bibitem[{{HI4PI Collaboration:} {et~al.}(2016){HI4PI Collaboration:}, {Ben
  Bekhti, N.}, {Fl\"oer, L.}, {Keller, R.}, {Kerp, J.}, {Lenz, D.}, {Winkel,
  B.}, {Bailin, J.}, {Calabretta, M. R.}, {Dedes, L.}, {Ford, H. A.}, {Gibson,
  B. K.}, {Haud, U.}, {Janowiecki, S.}, {Kalberla, P. M. W.}, {Lockman, F. J.},
  {McClure-Griffiths, N. M.}, {Murphy, T.}, {Nakanishi, H.}, {Pisano, D. J.},
  \& {Staveley-Smith, L.}}]{Bekhti_2016}
{HI4PI Collaboration:}, {Ben Bekhti, N.}, {Fl\"oer, L.}, {et~al.} 2016, A\&A,
  594, A116, hI4PI map

\bibitem[{{Liu} {et~al.}(2022){Liu}, {Bulbul}, {Ghirardini}, {Liu}, {Klein},
  {Clerc}, {{\"O}zsoy}, {Ramos-Ceja}, {Pacaud}, {Comparat}, {Okabe}, {Bahar},
  {Biffi}, {Brunner}, {Br{\"u}ggen}, {Buchner}, {Ider Chitham}, {Chiu},
  {Dolag}, {Gatuzz}, {Gonzalez}, {Hoang}, {Lamer}, {Merloni}, {Nandra},
  {Oguri}, {Ota}, {Predehl}, {Reiprich}, {Salvato}, {Schrabback}, {Sanders},
  {Seppi}, \& {Thibaud}}]{Liu_2022}
{Liu}, A., {Bulbul}, E., {Ghirardini}, V., {et~al.} 2022, \aap, 661, A2

\bibitem[{{Lodders}(2003)}]{Lodders_2003}
{Lodders}, K. 2003, \apj, 591, 1220

\bibitem[{{Markevitch} \& {Vikhlinin}(2007)}]{Markevitch2007}
{Markevitch}, M. \& {Vikhlinin}, A. 2007, Phys. Rep., 443

\bibitem[{Mei {et~al.}(2007)Mei, Blakeslee, Cote, Tonry, West, Ferrarese,
  Jordan, Peng, Anthony, \& Merritt}]{Mei_2007}
Mei, S., Blakeslee, J.~P., Cote, P., {et~al.} 2007, ApJ, 655, 144

\bibitem[{Merloni {et~al.}(2024)Merloni, Lamer, Teng, Ramos-Ceja, Brunner,
  Bulbul, Dennerl, Doroshenko, Freyberg, Friedrich, Gatuzz, \&
  Georgakakis}]{Merloni_2024}
Merloni, A., Lamer, G., Teng, L., {et~al.} 2024, A\&A, 682, A34

\bibitem[{Merloni {et~al.}(2012)Merloni, Predehl, \& the German~eROSITA
  Consortium}]{Merloni2012}
Merloni, A., Predehl, P., \& the German~eROSITA Consortium. 2012, A\& A

\bibitem[{{Million} {et~al.}(2011){Million}, {Werner}, {Simionescu}, \&
  {Allen}}]{Million_2011}
{Million}, E.~T., {Werner}, N., {Simionescu}, A., \& {Allen}, S.~W. 2011,
  \mnras, 418, 2744

\bibitem[{Mirakhor \& Walker(2021)}]{Mirakhor_2021}
Mirakhor, M.~S. \& Walker, S.~A. 2021, MNRAS, 506, 139

\bibitem[{{Nulsen} \& {B\"ohringer}(1995)}]{Nulsen_1995}
{Nulsen}, P.~E.~J. \& {B\"ohringer}, H. 1995, MNRAS, 274, 1093

\bibitem[{{Pacaud} {et~al.}(2006){Pacaud}, {Pierre}, {Refregier}, {Gueguen},
  {Starck}, {Valtchanov}, {Read}, {Altieri}, {Chiappetti}, {Gandhi}, {Garcet},
  {Gosset}, {Ponman}, \& {Surdej}}]{Pacaud_2006}
{Pacaud}, F., {Pierre}, M., {Refregier}, A., {et~al.} 2006, MNRAS, 372, 578

\bibitem[{{Piffaretti} {et~al.}(2011){Piffaretti}, {Arnaud}, {Pratt},
  {Pointecouteau}, \& {Melin}}]{Piffaretti_2011}
{Piffaretti}, R., {Arnaud}, M., {Pratt}, G.~W., {Pointecouteau}, E., \&
  {Melin}, J.~B. 2011, \aap, 534, A109

\bibitem[{{Predehl} {et~al.}(2021){Predehl}, {Andritschke}, {Arefiev},
  {Babyshkin}, {Batanov}, {Becker}, {B{\"o}hringer}, {Bogomolov}, {Boller},
  {Borm}, {Bornemann}, {Br{\"a}uninger}, {Br{\"u}ggen}, {Brunner}, {Brusa},
  {Bulbul}, {Buntov}, {Burwitz}, {Burkert}, {Clerc}, {Churazov}, {Coutinho},
  {Dauser}, {Dennerl}, {Doroshenko}, {Eder}, {Emberger}, {Eraerds},
  {Finoguenov}, {Freyberg}, {Friedrich}, {Friedrich}, {F{\"u}rmetz},
  {Georgakakis}, {Gilfanov}, {Granato}, {Grossberger}, {Gueguen}, {Gureev},
  {Haberl}, {H{\"a}lker}, {Hartner}, {Hasinger}, {Huber}, {Ji}, {Kienlin},
  {Kink}, {Korotkov}, {Kreykenbohm}, {Lamer}, {Lomakin}, {Lapshov}, {Liu},
  {Maitra}, {Meidinger}, {Menz}, {Merloni}, {Mernik}, {Mican}, {Mohr},
  {M{\"u}ller}, {Nandra}, {Nazarov}, {Pacaud}, {Pavlinsky}, {Perinati},
  {Pfeffermann}, {Pietschner}, {Ramos-Ceja}, {Rau}, {Reiffers}, {Reiprich},
  {Robrade}, {Salvato}, {Sanders}, {Santangelo}, {Sasaki}, {Scheuerle},
  {Schmid}, {Schmitt}, {Schwope}, {Shirshakov}, {Steinmetz}, {Stewart},
  {Str{\"u}der}, {Sunyaev}, {Tenzer}, {Tiedemann}, {Tr{\"u}mper}, {Voron},
  {Weber}, {Wilms}, \& {Yaroshenko}}]{Predehl_2021}
{Predehl}, P., {Andritschke}, R., {Arefiev}, V., {et~al.} 2021, \aap, 647, A1

\bibitem[{Predehl {et~al.}(2020)Predehl, Sunyaev, Becker, Brunner, Burenin,
  Bykov, Cherepashchuk, Chugai, Churazov, Doroshenko, Eismont, Freyberg,
  Gilfanov, Haberl, Khabibullin, Krivonos, Maitra, Medvedev, Merloni, Nandra,
  Nazarov, Pavlinsky, Ponti, Sanders, Sasaki, Sazonov, Strong, \&
  Wilms}]{Predehl_2020}
Predehl, P., Sunyaev, R.~A., Becker, W., {et~al.} 2020, Nature, 588, 227

\bibitem[{Randall {et~al.}(2008)Randall, Nulsen, Forman, Jones, Machacek,
  Murray, \& Maughan}]{Randall_2008}
Randall, S., Nulsen, P., Forman, W.~R., {et~al.} 2008, ApJ, 688, 208

\bibitem[{Reiprich {et~al.}(2013)Reiprich, Basu, \& Ettori}]{Reiprich2013}
Reiprich, T., Basu, K., \& Ettori, S. 2013, Space Science Reviews, 177,
  195–245

\bibitem[{Reiprich {et~al.}(2021)Reiprich, Veronica, Pacaud, Ramos-Ceja, Ota,
  Sanders, Kara, Erben, Klein, Erler, Kerp, Hoang, Brüggen, Marvil, Rudnick,
  Biffi, Dolag, Aschersleben, Basu, Brunner, Bulbul, Dennerl, Eckert, Freyberg,
  Gatuzz, Ghirardini, Käfer, Merloni, Migkas, Nandra, Predehl, Robrade,
  Salvato, Whelan, Diaz-Ocampo, Hernandez-Lang, Zenteno, Brown, Collier, Diego,
  Hopkins, Kapinska, Koribalski, Mroczkowski, Norris, O'Brien, \&
  Vardoulaki}]{Reiprich_2021}
Reiprich, T.~H., Veronica, A., Pacaud, F., {et~al.} 2021, Astronomy {\&}
  Astrophysics, 647, A2

\bibitem[{Sanders {et~al.}(2016)Sanders, Fabian, Taylor, Russell, Blundell,
  Canning, Hlavacek-Larrondo, Walker, \& Grimes}]{Sanders_2016}
Sanders, J.~S., Fabian, A.~C., Taylor, G.~B., {et~al.} 2016, MNRAS, 457, 82

\bibitem[{Schindler {et~al.}(1998)Schindler, Binggeli, \&
  B\"ohringer}]{Schindler_1998}
Schindler, S., Binggeli, B., \& B\"ohringer, H. 1998, Morphology of the Virgo
  Cluster: Gas versus Galaxies

\bibitem[{{Simionescu} {et~al.}(2007){Simionescu}, {B\"ohringer}, {Br\"uggen},
  \& {Finoguenov, A.}}]{Simionescu_2007}
{Simionescu}, {B\"ohringer}, H., {Br\"uggen}, M., \& {Finoguenov, A.} 2007,
  A\&A, 465, 749

\bibitem[{Simionescu {et~al.}(2008)Simionescu, Werner, Finoguenov, Böhringer,
  \& Brüggen}]{Simionescu_2008}
Simionescu, A., Werner, N., Finoguenov, A., Böhringer, H., \& Brüggen, M.
  2008, A \& A, 482, 97–112

\bibitem[{Simionescu {et~al.}(2010)Simionescu, Werner, Forman, Miller, Takei,
  Böhringer, Churazov, \& Nulsen}]{Simionescu_2010}
Simionescu, A., Werner, N., Forman, W.~R., {et~al.} 2010, MNRAS

\bibitem[{Simionescu {et~al.}(2017)Simionescu, Werner, Mantz, Allen, \&
  Urban}]{Simionescu_2017}
Simionescu, A., Werner, N., Mantz, A., Allen, S.~W., \& Urban, O. 2017, MNRAS,
  469, 1476

\bibitem[{SOC(2022)}]{xmm}
SOC, X.-N. 2022, Users Guide to the XMM-Newton Science Analysis System, ESA

\bibitem[{{Spasic} {et~al.}(2024){Spasic}, {Edler}, {Su}, {Br{\"u}ggen}, {de
  Gasperin}, {Pasini}, {Heesen}, {Simonte}, {Boselli}, {R{\"o}ttgering}, \&
  {Fossati}}]{Spasic_2024}
{Spasic}, A., {Edler}, H.~W., {Su}, Y., {et~al.} 2024, A\&A, Accepted for
  publication

\bibitem[{Su {et~al.}(2019)Su, Kraft, Nulsen, Jones, Maccarone, Mernier,
  Lovisari, Sheardown, Randall, Roediger, Fish, Forman, \& Churazov}]{Su_2019}
Su, Y., Kraft, R.~P., Nulsen, P. E.~J., {et~al.} 2019, The Astronomical
  Journal, 158, 6

\bibitem[{Sunyaev {et~al.}(2021)Sunyaev, Arefiev, Babyshkin, Bogomolov,
  Borisov, Buntov, Brunner, Burenin, Churazov, Coutinho, Eder, Eismont,
  Freyberg, Gilfanov, Gureyev, Hasinger, Khabibullin, Kolmykov, Komovkin,
  Krivonos, Lapshov, Levin, Lomakin, Lutovinov, Medvedev, Merloni, Mernik,
  Mikhailov, Molodtsov, Mzhelsky, Müller, Nandra, Nazarov, Pavlinsky,
  Poghodin, Predehl, Robrade, Sazonov, Scheuerle, Shirshakov, Tkachenko, \&
  Voron}]{Sunyaev_2021}
Sunyaev, R., Arefiev, V., Babyshkin, V., {et~al.} 2021, A \& A, 656, A132

\bibitem[{Tonry {et~al.}(2001)Tonry, Dressler, Blakeslee, Ajhar, Fletcher,
  Luppino, Metzger, \& Moore}]{Tonry_2001}
Tonry, J.~L., Dressler, A., Blakeslee, J.~P., {et~al.} 2001, ApJ, 546, 681

\bibitem[{Urban {et~al.}(2011)Urban, Werner, Simionescu, Allen, \&
  Böhringer}]{Urban2011}
Urban, O., Werner, N., Simionescu, A., Allen, S.~W., \& Böhringer, H. 2011,
  MNRAS, 414, 2101–2111

\bibitem[{van Weeren {et~al.}(2019)van Weeren, de~Gasperin, Akamatsu,
  Br{\"u}ggen, Feretti, Kang, Stroe, \& Zandanel}]{vanWeeren_2019}
van Weeren, R.~J., de~Gasperin, F., Akamatsu, H., {et~al.} 2019, Space Science
  Reviews, 215, 16

\bibitem[{{Vazza} {et~al.}(2013){Vazza}, {Eckert}, {Simionescu}, {Br{\"u}ggen},
  \& {Ettori}}]{Vazza_2013}
{Vazza}, F., {Eckert}, D., {Simionescu}, A., {Br{\"u}ggen}, M., \& {Ettori}, S.
  2013, \mnras, 429, 799

\bibitem[{Virtanen {et~al.}(2020)Virtanen, Gommers, Oliphant, Haberland, Reddy,
  Cournapeau, Burovski, Peterson, Weckesser, Bright, {van der Walt}, Brett,
  Wilson, Millman, Mayorov, Nelson, Jones, Kern, Larson, Carey, Polat, Feng,
  Moore, {VanderPlas}, Laxalde, Perktold, Cimrman, Henriksen, Quintero, Harris,
  Archibald, Ribeiro, Pedregosa, {van Mulbregt}, \& {SciPy 1.0
  Contributors}}]{scipy}
Virtanen, P., Gommers, R., Oliphant, T.~E., {et~al.} 2020, Nature Methods, 17,
  261

\bibitem[{Voit(2005)}]{Voit_2005}
Voit, G.~M. 2005, Reviews of Modern Physics, 77, 207

\bibitem[{Walker {et~al.}(2019)Walker, Simionescu, Nagai, Okabe, Eckert,
  Mroczkowski, Akamatsu, Ettori, \& Ghirardini}]{Walker_2019}
Walker, S., Simionescu, A., Nagai, D., {et~al.} 2019, Space Science Reviews,
  215

\bibitem[{Walker {et~al.}(2016)Walker, Sanders, \& Fabian}]{Walker_2016}
Walker, S.~A., Sanders, J.~S., \& Fabian, A.~C. 2016, MNRAS, 461, 684

\bibitem[{{Werner} {et~al.}(2006){Werner}, {B{\"o}hringer}, {Kaastra}, {de
  Plaa}, {Simionescu}, \& {Vink}}]{Werner_2006}
{Werner}, N., {B{\"o}hringer}, H., {Kaastra}, J.~S., {et~al.} 2006, \aap, 459,
  353

\bibitem[{{Werner} {et~al.}(2016){Werner}, {ZuHone}, {Zhuravleva}, {Ichinohe},
  {Simionescu}, {Allen}, {Markevitch}, {Fabian}, {Keshet}, {Roediger},
  {Ruszkowski}, \& {Sanders}}]{Werner_2016}
{Werner}, N., {ZuHone}, J.~A., {Zhuravleva}, I., {et~al.} 2016, \mnras, 455,
  846

\bibitem[{{Yeung} {et~al.}(2023){Yeung}, {Freyberg}, {Ponti}, {Dennerl, K.},
  {Sasaki, M.}, \& {Strong, A.}}]{Yeung_2023}
{Yeung}, M. C.~H., {Freyberg}, M.~J., {Ponti}, G., {et~al.} 2023, A\&A, 676, A3

\bibitem[{Young {et~al.}(2002)Young, Wilson, \& Mundell}]{Young_2002}
Young, A.~J., Wilson, A.~S., \& Mundell, C.~G. 2002, ApJ, 579, 560

\bibitem[{{Zhuravleva} {et~al.}(2013){Zhuravleva}, {Churazov}, {Kravtsov},
  {Lau}, {Nagai}, \& {Sunyaev}}]{Zhuravleva_2013}
{Zhuravleva}, I., {Churazov}, E., {Kravtsov}, A., {et~al.} 2013, \mnras, 428,
  3274

\bibitem[{{Zinger} {et~al.}(2016){Zinger}, {Dekel}, {Birnboim}, {Kravtsov}, \&
  {Nagai}}]{Zinger_2016}
{Zinger}, E., {Dekel}, A., {Birnboim}, Y., {Kravtsov}, A., \& {Nagai}, D. 2016,
  \mnras, 461, 412

\bibitem[{Zinger {et~al.}(2018)Zinger, Dekel, Birnboim, Nagai, Lau, \&
  Kravtsov}]{Zinger_2018}
Zinger, E., Dekel, A., Birnboim, Y., {et~al.} 2018, MNRAS, 476, 56

\end{thebibliography}
%
\begin{appendix}

\section{Observations and data reduction products} \label{app:Observations}

\begin{table}[H]
	\centering
 	\caption{Important TM-specific values for the PIB subtraction.}
	\begin{tabular}{ccc}
            \hline
		TM & $R$ \tablefootmark{a} & $H_\mathrm{obs}$ \tablefootmark{b}\\
            \hline
		1 & 1.05 & 203152 \\
		2 & 1.12 & 185851 \\
		3 & 1.08 & 199857 \\
		4 & 1.01 & 213538 \\
		5 & 0.69 & 252434 \\
		6 & 1.04 & 198713 \\
		7 & 0.66 & 255768 \\
            \hline
	\end{tabular}
        \vspace{0.2 cm}
    \tablefoot{
    \tablefoottext{a}{$R$ is the ratio of the number of counts in the Filter-Wheel-Closed (FWC) data in the soft band (0.3-2.0\,keV) to the hard band (6.7-9.0\,keV).}
    \tablefoottext{b}{$H_\mathrm{obs}$ is the hard band counts.}}
	\label{tab:PIB}
\end{table}

\begin{figure}[H]
	\centering
	\includegraphics[width=0.6\linewidth]{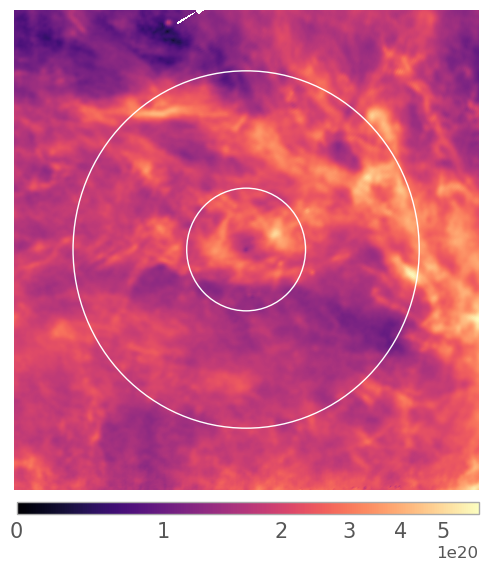}
	\caption{$N_{HI}$ map from HI4PI trimmed to match the area of the sky analyzed in this work. The color bar is in units of cm$^{-2}$. $r_{200}$ and $3r_{200}$ are plotted.}
	\label{fig:nh_map}
\end{figure}

\begin{figure}[H]
	\centering
	\includegraphics[width=0.95\linewidth]{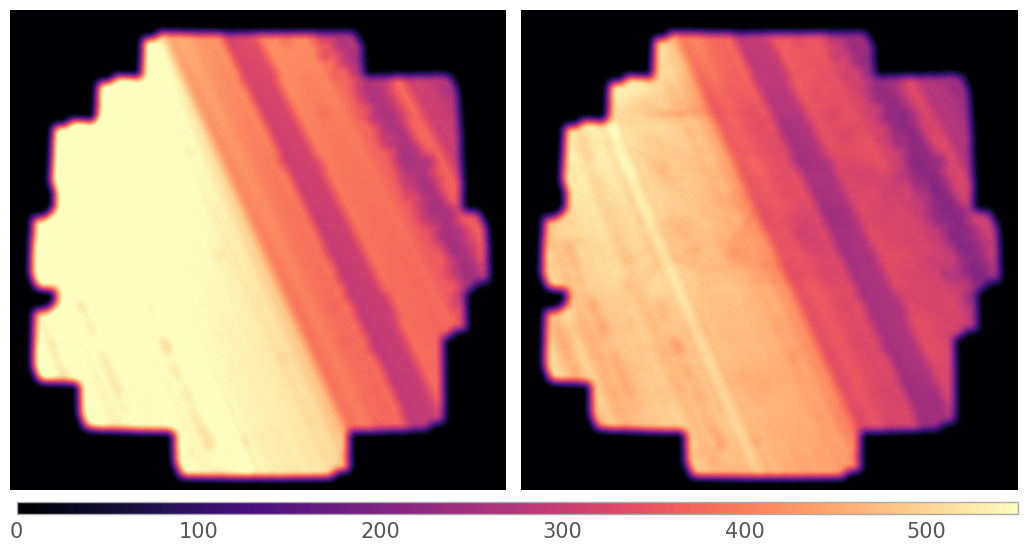}
	\caption{TM\,0 exposure maps from various stages in the data reduction. Left: Following \texttt{flaregti} application, but before the application of any additional corrections. Right: Following flare filtering, absorption correction, and exposure correction steps.}
	\label{fig:exp_maps}
\end{figure}

\begin{table}[H]
	\centering
     \caption{Simulated spectra values from the absorption correction step.}
	\begin{tabular}{cc}
            \hline
		LHB Temp. (keV) & 0.099\\
            MWH Temp. (keV) & 0.225\\
            PL Photon Index & 1.4\\
            LHB Norm. ($\mathrm{cm}^{-5}/\mathrm{deg}^2$) & 0.0019\\
            MWH Norm ($\mathrm{cm}^{-5}/\mathrm{deg}^2$) & 0.0041\\
            PL Norm ($\mathrm{photons}/\mathrm{keV}/\mathrm{cm}^2/\mathrm{s}/\mathrm{deg}^2$) & 0.0036\\
            \hline
	\end{tabular}
        \vspace{0.2 cm}
	\label{tab:nh}
\end{table}
\FloatBarrier

\begin{table*}
\centering
\caption{All eROSITA observations of the Virgo Cluster used in this work.}
\begin{tabular}{cc||cc}
  \hline
  \textbf{Tile} & \textbf{Excluded eRASS: TM} & \textbf{Tile} & \textbf{Excluded eRASS: TM} \\
  \hline
  190066 & 3: 4, 5, 7; 4: 4; 5 & 190069 & 3: 4, 5, 7; 4: 4; 5\\
  186066 & 3: 4, 5, 7; 4: 4; 5 & 186069 & 3: 4, 5, 7; 4: 4; 5\\
  183066 & 3: 4, 5, 7; 4: 4; 5 & 183069 & 3: 4, 5, 7; 4: 4; 5\\
  180069 & 3: 4, 5, 7; 4: 4; 5 & 196072 & 4: 4\\
  180072 & 4: 4; 5 & 198075 & 4: 4\\
  195075 & 4: 4 & 180075 & 3: 4, 5, 7; 4: 4; 5\\
  177075 & 3: 4; 4: 4; 5 & 200078 & 4: 4\\
  197078 & 4: 4 & 198081 & 4: 4\\
  195081 & 4: 4 & 200084 & 4: 4\\
  197084 & 4: 4 & 194084 & 4: 4\\
  194087 & 4: 4 & 191087 & 4: 4\\
  188087 & 4: 4 & 185087 & 4: 4\\
  191090 & 4: 4 & 188090 & 4: 4\\
  185090 & 4: 4 & 182087 & 3: 4, 5, 7; 4: 4; 5\\
  179087 & 3: 4, 5, 7; 4: 4; 5 & 182084 & 3: 4, 5, 7; 4: 4; 5\\
  179084 & 3: 4, 5, 7; 4: 4; 5 & 180081 & 3: 4, 5, 7; 4: 4; 5\\
  177081 & 3: 4, 5, 7; 4: 4; 5 & 178078 & 3: 4, 5, 7; 4: 4; 5\\
  175078 & 3: 4; 4: 4; 5 & 193072 & 4: 4\\
  189072 & 4: 4 & 186072 & 3: 4, 5, 7; 4: 4; 5\\
  183072 & 3: 4, 5, 7; 4: 4; 5 & 192075 & 4: 4\\
  194078 & 4: 4 & 192081 & 4: 4\\
  191084 & 4: 4 & 188084 & 4: 4\\
  185084 & 4: 4 & 183081 & 3: 4, 5, 7; 4: 4; 5\\
  191078 & 4: 4 & 189081 & 4: 4\\
  189075 & 4: 4 & 188078 & 4: 4\\
  186081 & 4: 4 & 186075 & 3: 4, 5, 7; 4: 4; 5\\
  185078 & 3: 4, 5, 7; 4: 4; 5 & 183075 & 3: 4, 5, 7; 4: 4; 5\\
  182078 & 3: 4, 5, 7; 4: 4; 5 & 177072 & 3: 4; 4: 4; 5\\
  193069 & 4: 4 & 193066 & 4: 4\\
  197087 & 4: 4 & 199072 & 4: 4\\
  177069 & 4: 4; 5 & 196069 & 4: 4\\
  \hline
\end{tabular}
\label{tab:obs}
\tablefoot{TM\,4 of eRASS\,4 was excluded from every tile due to micrometeorite impacts\footnotemark. TMs 4, 5, and 7 were unavailable for several tiles in eRASS\,3. Only a portion of eRASS\,5 was completed.}
\end{table*}
\FloatBarrier
\footnotetext{This was discussed in internal communications at the eROSITA consortium meeting in January 2022.}
\clearpage

\section{Surface brightness analysis products} \label{app:sb}

\begin{figure}[h!]
	\centering
\includegraphics[width=0.8\linewidth]{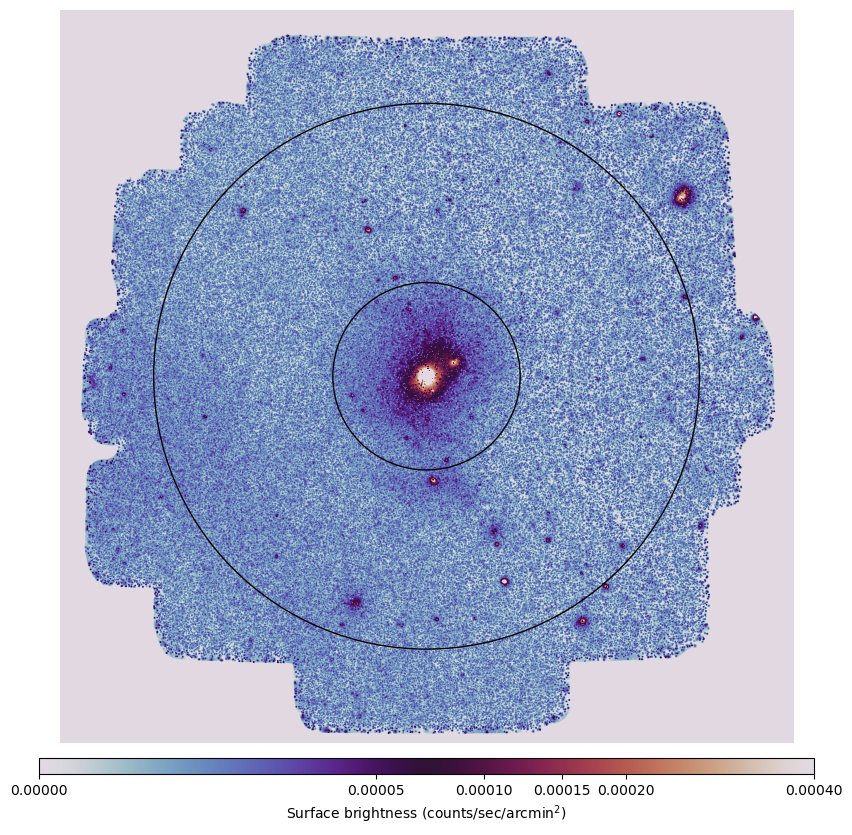}
	\caption{1.0-2.3\,keV (blue band) wavelet filtered image with point sources removed. $r_{200}$ and $3r_{200}$ are plotted. Emission from the eROSITA bubble is faintly visible only beyond $r_{200}$.}
	\label{fig:blue_img}
\end{figure}

\begin{figure}[h!]
	\centering
	\includegraphics[width=0.78\linewidth]{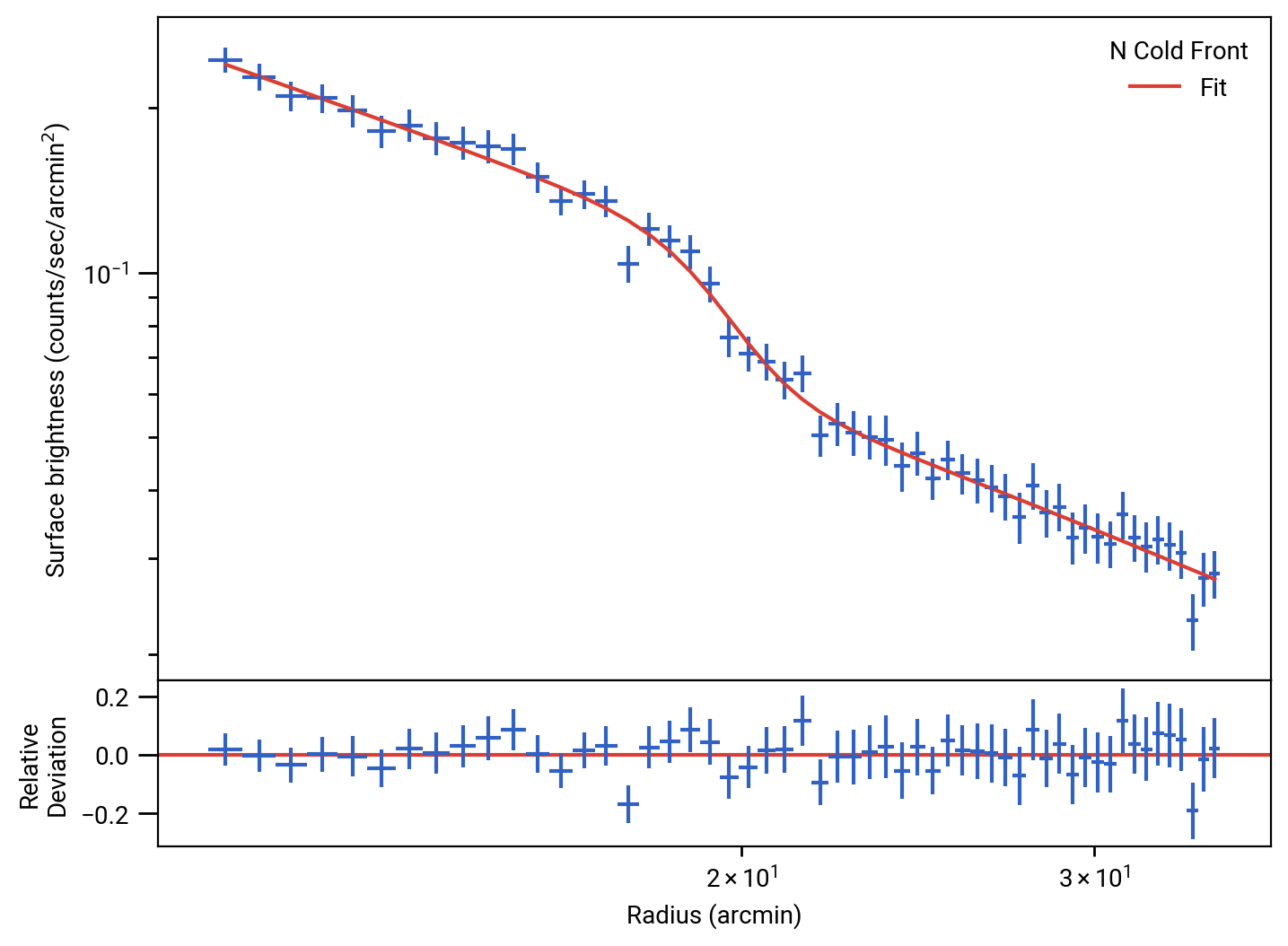}
	\caption{0.3-2.0\,keV eROSITA surface brightness profile for the northern cold front at $90$\,kpc fitted to a single $\beta$-model with a discontinuity.}
	\label{fig:northern_sector}
\end{figure}

\begin{figure}[h!]
	\centering
	\includegraphics[width=0.78\linewidth]{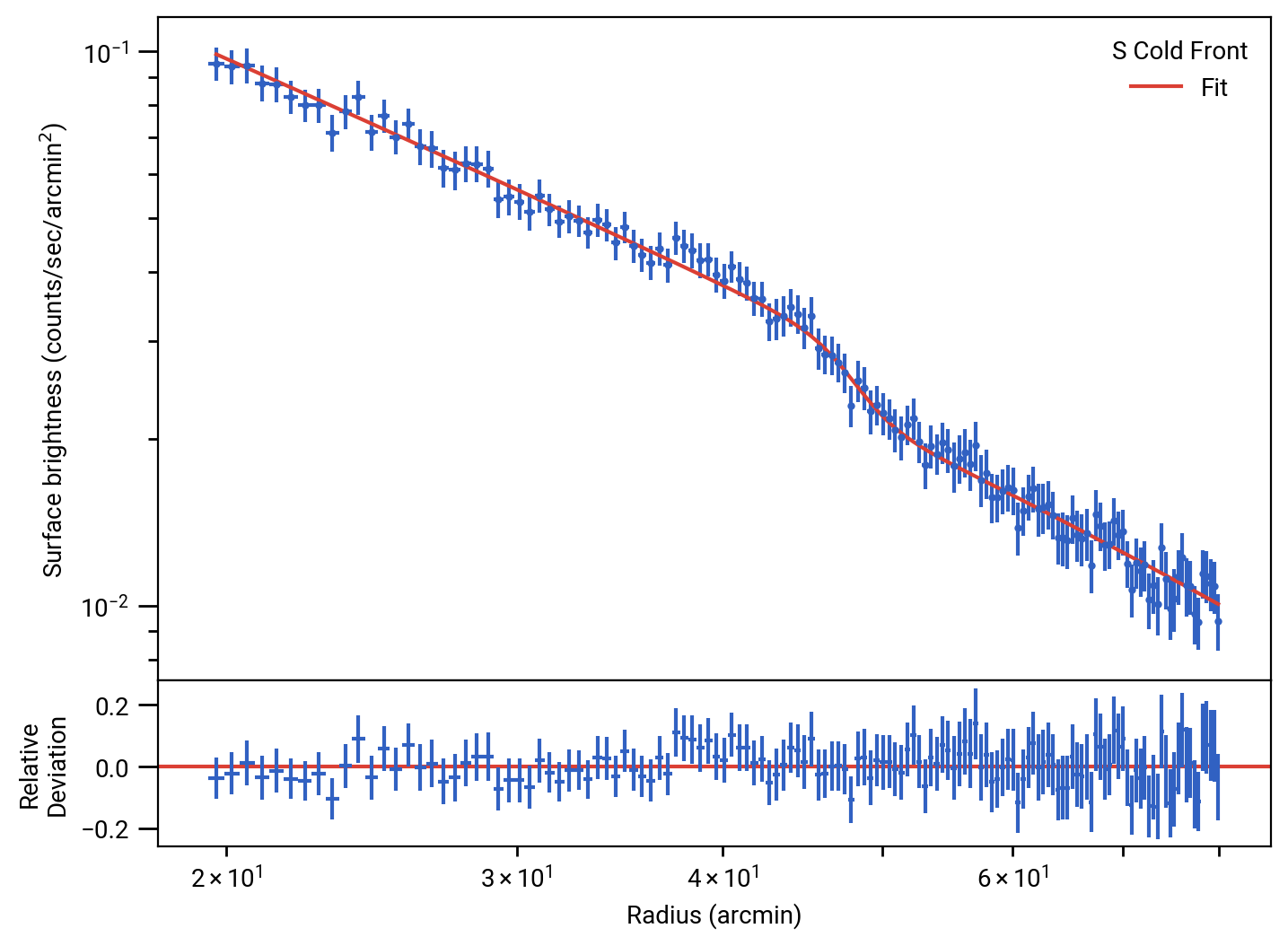}
	\caption{0.3-2.0\,keV eROSITA surface brightness profile for the southern cold front at $\sim\ 220$\,kpc fitted to a single $\beta$-model with a discontinuity.}
	\label{fig:southern_sector}
\end{figure}

\begin{figure}
	\centering
	\includegraphics[width=0.8\linewidth]{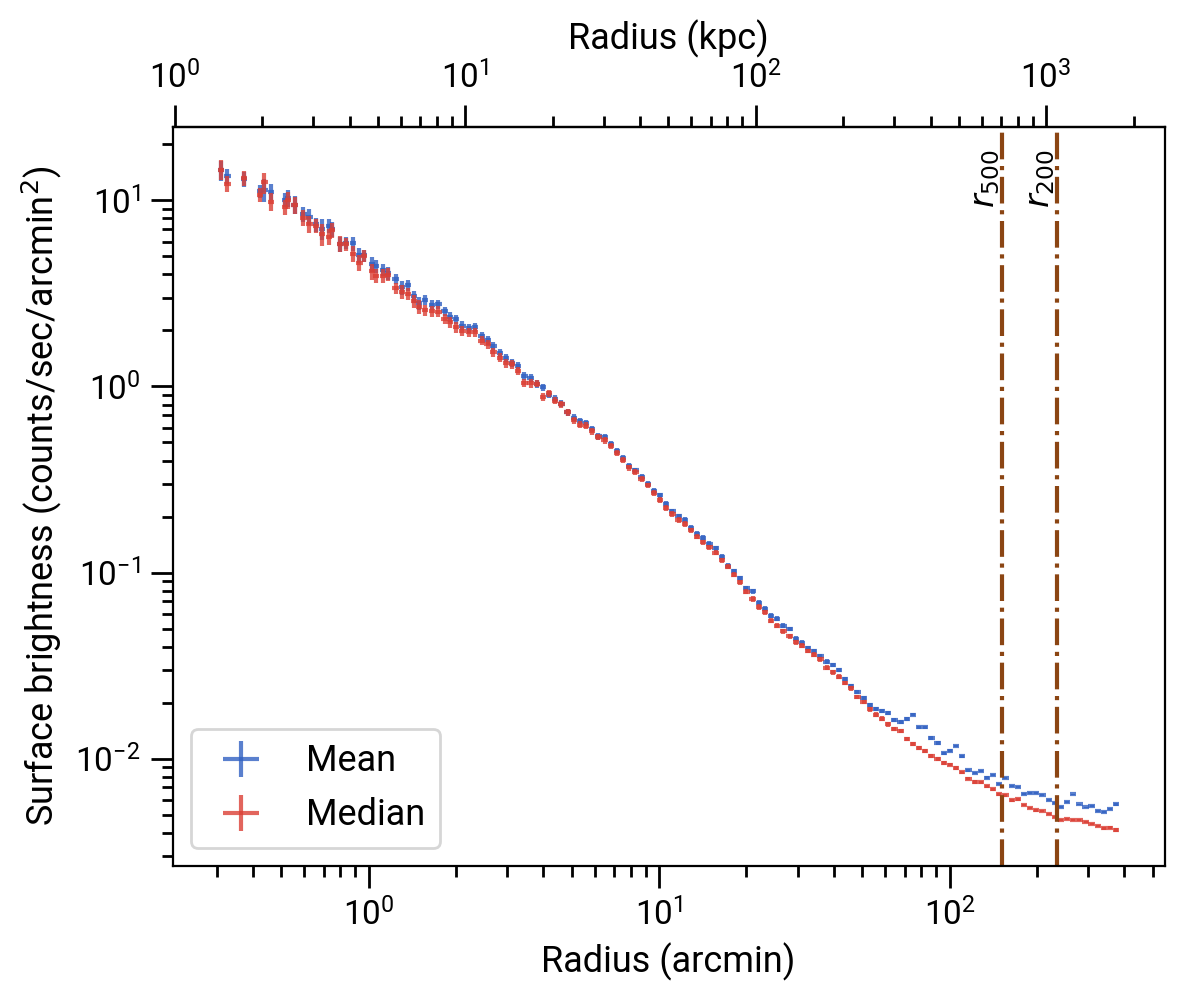}
	\caption{Surface brightness profiles in the $0.3-2.0$\,keV band obtained by performing the azimuthal mean (blue) and azimuthal median (red).}
	\label{fig:median_sb}
\end{figure}

\begin{figure}
	\centering
	\includegraphics[width=0.8\linewidth]{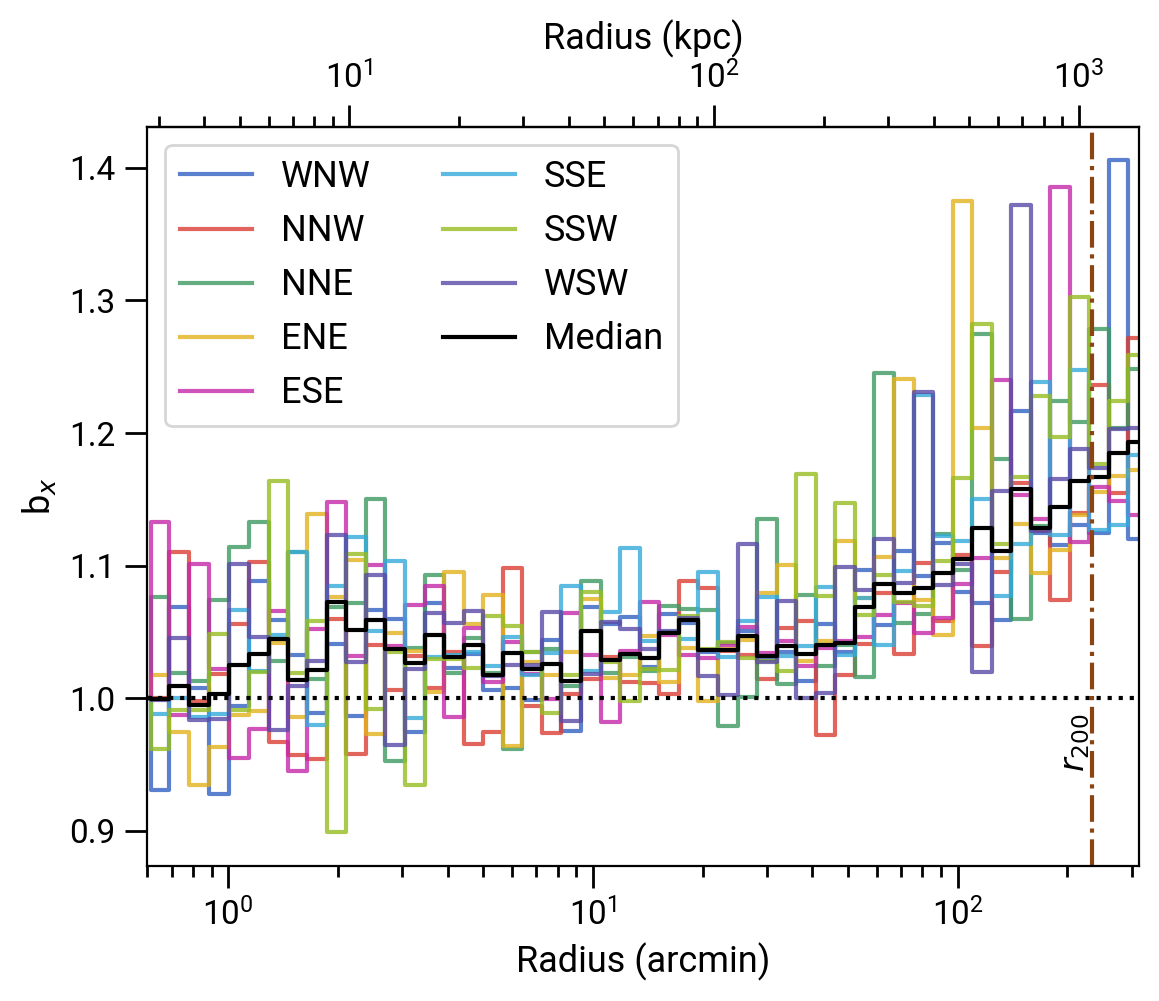}
	\caption{Emissivity bias, $b_X$, for eight sectors as a function of distance from cluster center after removal of the extended gas halos around M86, M58, and M49. The sectors are color-coded according to Fig.~\ref{fig:SB_8profs_key}. The median of all sectors is shown in black.}
	\label{fig:emiss_bias8}
\end{figure}

\begin{figure}
	\centering
	\includegraphics[width=0.8\linewidth]{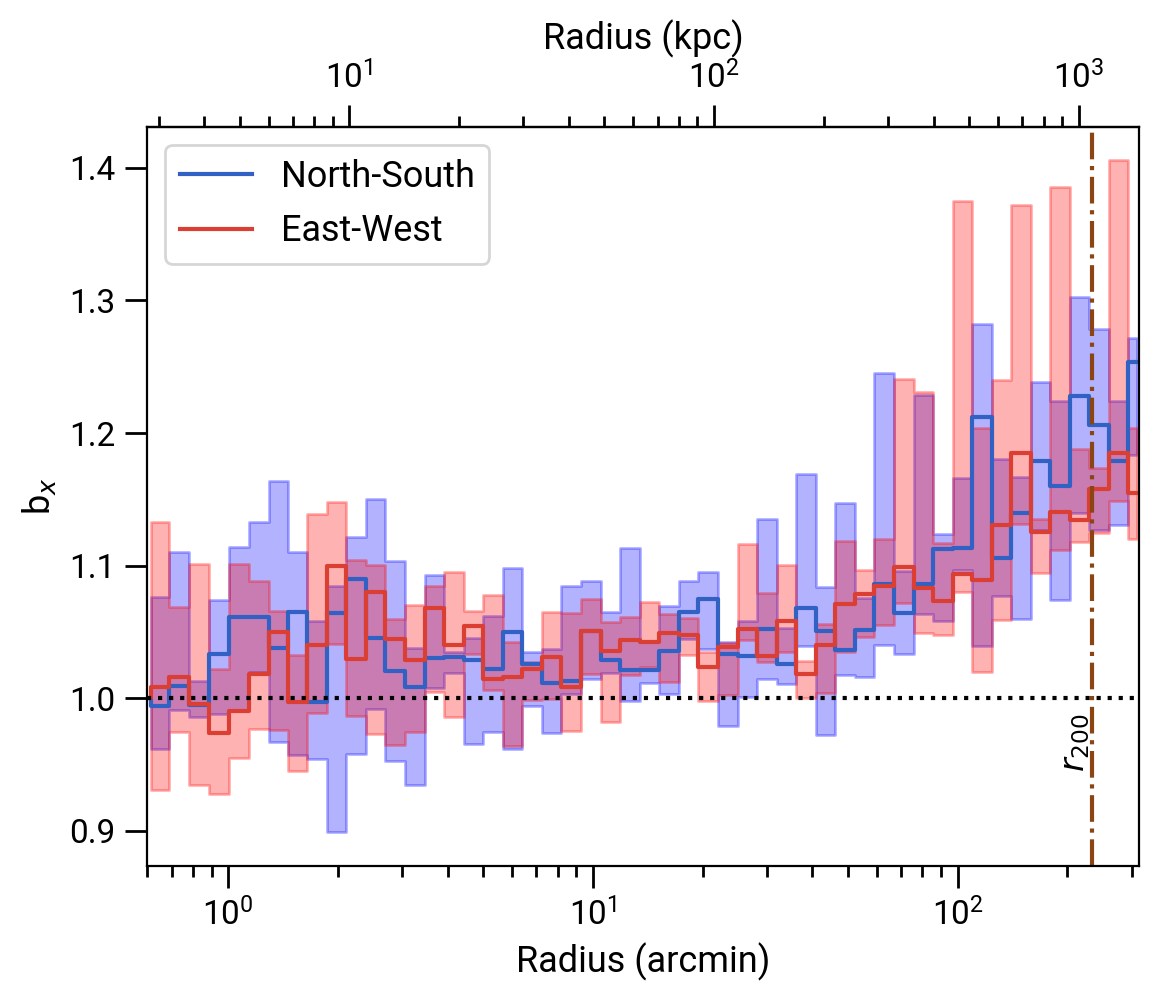}
	\caption{Emissivity bias, $b_X$, of the north-south axis and east-west axis of the cluster. Shaded regions show the range of values in each bin, not uncertainties.}
	\label{fig:emiss_bias_NS}
\end{figure}

\begin{figure*}
	\centering
        \includegraphics[width=0.94\linewidth]{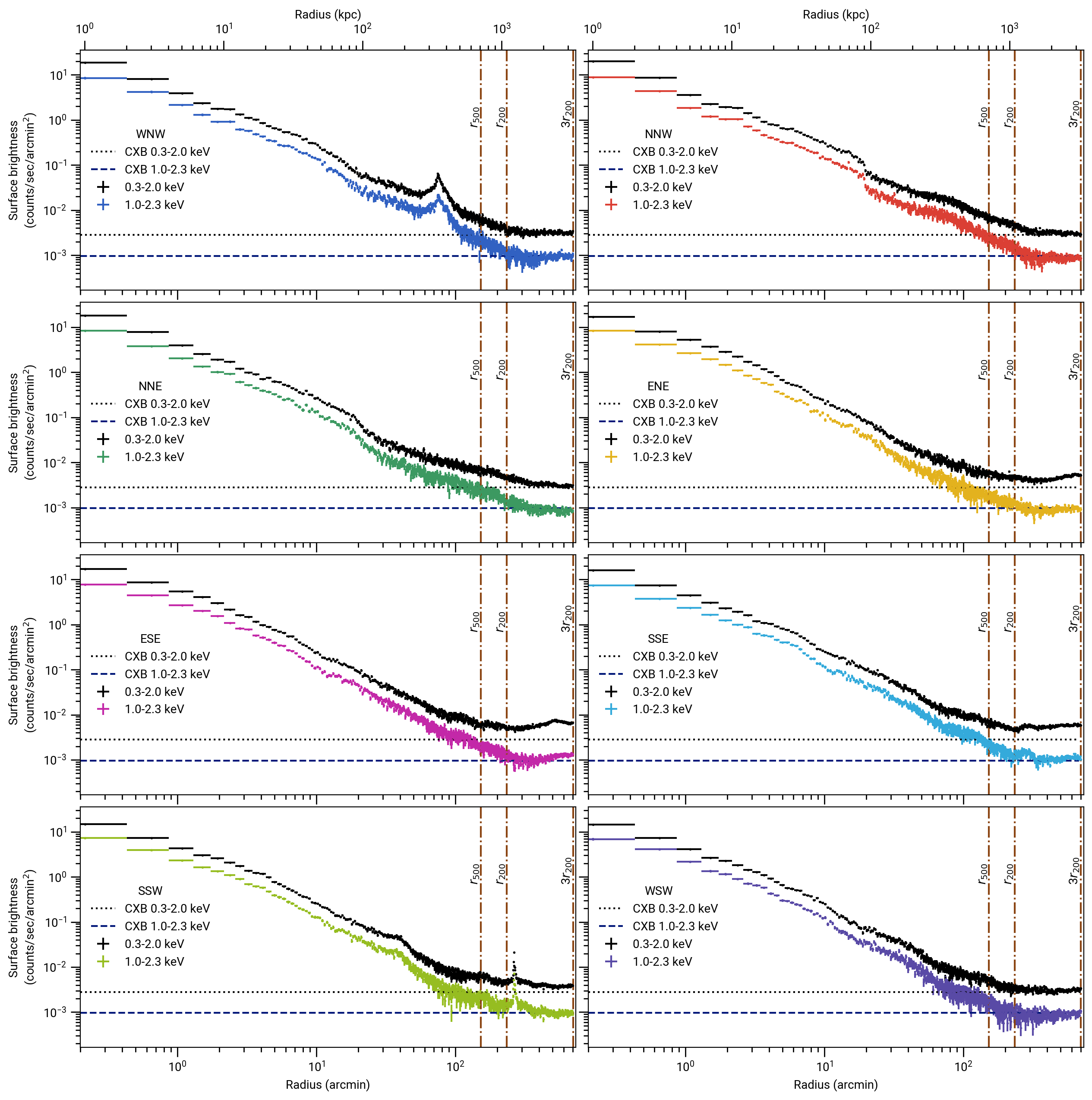}
	\caption{Eight surface brightness profiles in the $1.0-2.3$\,keV (blue) energy band, one for each sector identified in Fig.~\ref{fig:SB_8profs_key}. The $0.3-2.0$\,keV band for any given profile is plotted in black for comparison. The $0.3-2.0$\,keV CXB measured by eROSITA in the northwest of the cluster is shown as a dotted black line on each plot. The blue band CXB estimate is taken from the same regions as the $0.3-2.0$\,keV CXB, and shown as a dashed navy blue line. A selection of characteristic radii are shown as dash-dotted vertical brown lines. }
	\label{fig:SB_8profs_blue}
\end{figure*}

\end{appendix}

\end{document}